\documentstyle[epsfig]{mn}
 at 10truept
\title
     [Putting the Precision in Precision Cosmology]
{\vglue-3.0truecm \centerline{\it\small 
For submission to Monthly Notices}
\vglue 2.5truecm
 Putting the Precision in Precision Cosmology:\\
  How accurate should your data covariance matrix be? 
\author
     [ Andy Taylor, Benjamin Joachimi, Thomas Kitching]
     { Andy Taylor$^1$\thanks{ant@roe.ac.uk}, Benjamin Joachimi$^1$, Thomas Kitching$^{1,2}$\\
  1.	Scottish Universities Physics Alliance (SUPA), 
     Institute for Astronomy,
     School of Physics and Astronomy,\\
     University of Edinburgh,
     Royal Observatory,
     Blackford Hill,
     Edinburgh, EH9 3HJ,
     U.K. \\
     2.  Mullard Space Science Laboratory, University College London, Holmbury St. Mary, Surrey, RH5 6NT, U.K.
     } }
\newcommand{\be}{\begin{equation}}
\newcommand{\ee}{\end{equation}}
\newcommand{\ba}{\begin{eqnarray}}
\newcommand{\ea}{\end{eqnarray}}
\newcommand{\nn}{\nonumber \\}

\newcommand{\r}{\mbox{\boldmath $r$}}

\newcommand{\thetab}{\mbox{\boldmath $\theta$}}

\newcommand{\de}{\partial}

\newcommand{\lgl}{\langle}
\newcommand{\rgl}{\rangle}

\newcommand{\Tr}{\mbox{\rm Tr}}
\newcommand{\lb}{\mbox{\boldmath $\ell$}}

\newcommand{\x}{\mbox{\boldmath $x$}}
\newcommand{\k}{\mbox{\boldmath $k$}}

\newcommand{\D}{\mbox{\boldmath $D$}}
\newcommand{\C}{\mbox{\boldmath $C$}}
\newcommand{\Phib}{\mbox{\boldmath $\Phi$}}

\newcommand{\I}{\mbox{\boldmath $I$}}
\newcommand{\A}{\mbox{\boldmath $A$}}

\newcommand{\V}{\mbox{\boldmath $V$}}

\newcommand{\half}{\frac{1}{2}}

\newcommand{\Mb}{\mbox{\boldmath $M$}}

\newcommand{\W}{\mbox{\boldmath $W$}}

\newcommand{\calF}{{\cal F}}

\newcommand{\mub}{\mbox{\boldmath $\mu$}}

\newcommand{\calL}{{\cal L}}
\newcommand{\ep}{\varepsilon}
\newcommand{\Y}{\mbox{\boldmath $Y$}}
\newcommand{\X}{\mbox{\boldmath $X$}}
\newcommand{\Vb}{\mbox{\boldmath $V$}}
\newcommand{\Sigmab}{\mbox{\boldmath $\Sigma$}}
\newcommand{\Jb}{\mbox{\boldmath $J$}}
\newcommand{\Thetab}{\mbox{\boldmath $\Theta$}}
\newcommand{\U}{\mbox{\boldmath $U$}}
\newcommand{\G}{\mbox{\boldmath $G$}}
\newcommand{\y}{\mbox{\boldmath $y$}}
\newcommand{\B}{\mbox{\boldmath $B$}}
\newcommand{\Z}{\mbox{\boldmath $Z$}}
\newcommand{\calM}{{\cal M}}
\newcommand{\Tb}{\mbox{\boldmath $T$}}
\newcommand{\Psib}{\mbox{\boldmath $\Psi$}}
\newcommand{\calT}{{\cal T}}
\newcommand{\calTb}{\mbox{\boldmath $\calT$}}

\begin{document}

\maketitle
\begin{abstract}
Cosmological parameter estimation  requires that the likelihood function of the data is accurately known. 
Assuming  that cosmological large-scale structure power spectra data
 are multivariate Gaussian-distributed, we show the accuracy of parameter estimation   is  limited by the accuracy of the  inverse data covariance matrix  -- the {\em precision matrix}. 
If the data covariance and precision matrices are estimated by sampling independent realisations of the data, their statistical properties  are described by the Wishart and Inverse-Wishart distributions, respectively. 
Independent of any details of the survey, we show that the fractional error on a parameter variance, or a Figure-of-Merit, is equal to the fractional variance of the precision matrix. In addition, for the only unbiased estimator of the precision matrix,
we find that the fractional accuracy of the parameter error depends only on the difference between the number of independent realisations and the number of data points, and so can easily diverge.
For a $5\%$ error on a parameter error  and $N_D  \ll 10^2$ data-points, a minimum of 200 realisations of the survey are needed, with $10\%$ accuracy in the data covariance.
 If the number of data-points $N_D \gg 10^2$ we need  $N_S > N_D$ realisations and a fractional accuracy of $<\sqrt{2/N_D}$ in the data covariance. As the number of power spectra data points grows to $N_D>10^4$ -- $10^6$ this approach will be problematic. 
We discuss possible ways to relax these conditions: improved theoretical modelling; shrinkage methods; data-compression;
simulation and data resampling methods.
\end{abstract}



\section{Introduction}

A central part of modern cosmology is the measurement of the parameters that characterise cosmological models of the Universe. These can be the set that constitutes the Standard Cosmological Model $(\Omega_m,\Omega_b,\Omega_\Lambda, H_0, \sigma_8, n_s, \tau)$, or an extended set that characterise, for example, more complex dark energy models (see e.g., Copeland, Sami \& Tsujikawa, 2006, Amendola et al., 2012, for reviews), deviations from Einstein gravity (e.g., Clifton, Ferreira, Padillo, Skordis, 2012; Amendola et al., 2012 for recent reviews), more detail about the inflationary epoch (e.g., Amendola et al, 2012), isocurvature density and velocity modes (e.g., Bucher, Moodley \& Turok, 2001), or massive neutrinos and their abundance (e.g., Bird, Viel \& Haehnelt, 2012, and references therein). Furthermore, if we want to differentiate between theoretical models in a Bayesian framework, as well as estimate their parameter value, we also need to accurately  integrate over the model parameter-space (e.g., Trotta, 2007; Liddle, Mukherjee, Parkinson, 2006; Taylor \& Kitching, 2010). 

To carry these tasks out we need both accurate theoretical predictions of the physical properties  of the model to compare to the data, and sufficiently accurate models of their statistical properties. Ideally, we would like to be able to accurately predict the full multivariate probability distribution of the data for each model.  If, as is commonly assumed, the data can be modelled as a multivariate Gaussian distribution, all of the statistical properties of the model reside in the mean and covariance of the model. Attention has been focussed on the accuracy of the predictions of the mean value -- e.g., the model power spectra  -- and the effect of biases or errors in the mean  (e.g., Huterer \& Takada, 2005;  Huterer, et al., 2006; Taylor et al., 2007). But to 
fully specify  the distribution of the data we also need accurate predictions of the data covariance matrix and the inverse of the data covariance -- the {\em precision matrix}.  

If we assume that the mean is well-known,  the accuracy of the probability distribution of the data, and hence the likelihood function in parameter-space, is  determined by  the accuracy of the precision matrix. 
However, as yet there is no unique approach to estimating the data covariance matrix since this may depend on the details of what is known about the data, and even less attention paid to the estimation of the precision matrix.

The data covariance matrix can be estimated in a number of ways: direct calculation of a theoretical model; estimate the sample covariance from an ensemble of simulations of the data;  or estimate the sample covariance matrix from the data itself. 
If we know the data covariance from the theoretical model accurately enough, there is no statistical uncertainty, and the precision matrix can be accurately estimated.
 But if the data covariance matrix must be sampled from an ensemble of simulations, or the data itself, there will be statistical uncertainty in the sample covariance. If we assume the underlying data is Gaussian-distributed and the samples are independent and drawn from the same distribution, the 
 probability distribution of the sample covariance matrix is known, and was first derived by Wishart (1928; see also e.g., Press, 1982).
To fully specify the model distribution of the data we also require the precision matrix. The distribution of the precision matrix, the Inverse-Wishart distribution (e.g., Press, 1982), has significantly different properties from distribution of the covariance matrix. The Wishart distributions has previously been discussed in cosmology as the distribution of Cosmic Microwave Background (CMB) temperature and polarisation power spectra (Percival \& Brown, 2006), while the Inverse-Wishart distribution has been used as a prior for Bayesian estimates of the CMB temperature and polarisation power spectra (Eriksen \& Wehus, 2009), for Gibbs sampling (Larson et al., 2007), and to test Pseudo-Cl methods (Hamimeche \& Lewis, 2009).

In this paper we develop a new framework to estimate the statistical error on the data covariance and precision matrix (Section \ref{sec:cov_prec}). We illustrate these effects on simulated data (Section \ref{sec:example}) and discuss the implications for imminent and future large-scale structure surveys in cosmology. These effects are propagated into the accuracy of parameter errors, and  the parameter covariance around the peak of the likelihood surface (Section \ref{sec:paracovmat}). Since many experiments use the 2-parameter Figure-of-Merit (FoM) as a target measure for survey design, we also discuss the accuracy of  an arbitrary FoM  (Section \ref{sec:FoM}). Given a prescribed accuracy for the parameter covariance matrix, or a FoM, we show how accurate the precision matrix and data covariance matrix  must be. Finally, we discuss ways in which we avoid these bounds by improved theoretical modelling of the data covariance, rapid simulation production, or using data compression and shrinkage methods (Section \ref{sec:8}). 
We begin by reviewing parameter estimation and the role of the precision matrix.

\section{Parameter Estimation}

To begin with we shall assume that the cosmological parameters, $\thetab$, being measured are estimated from maximising a posterior parameter distribution, $p(\thetab|\D,\calM)$, given a dataset, $\D$, and some theoretical model, $\calM$  (see, e.g., Sivia, 1996). From Bayes Theorem,
 \be
 	p(\thetab|\D,\calM) = \frac{L(\D|\thetab,\calM) \pi(\thetab|\calM)}{E(\D|\calM)},
 \ee
we can determine the posterior parameter distribution from the likelihood function for the data, $L(\D|\thetab, \calM)$,  predicted by the model, a prior, $\pi(\thetab|\calM)$, which is the probability distribution of the parameters before the data is analysed, and normalised by the evidence, $E(\D|\calM)$, which marginalises over the likelihood and prior in parameter-space. If we restrict our study to parameter estimation for a given model, we can ignore this term. We shall assume the prior on the parameters is flat. 

 If we model the data distribution as a multivariate Gaussian, then the likelihood function can be written
%
 \be
 	L(\D|\mub,\Mb,\calM) = \frac{1}{(2 \pi)^{N_D/2} \sqrt{|\Mb|}} \exp - \half \Tr  \,  \W \Psib,
 \ee
where 
 \be
 	\W =   \Delta\! \D \Delta \! \D^t,
 \ee
 a superscript, $t$, indicates a transpose, 
  \be
	  \Delta \! \D = \D - \lgl \D \rgl
  \ee
  is the variation in the data-vector,  $\mub=\lgl \D \rgl$ is  the mean of the data
and  $N_D$ is the length of the data-vector.  
  The data covariance matrix is given by
 \be
	 	\Mb = \lgl \W \rgl = \lgl \Delta\! \D \Delta \! \D^t\rgl  .
 \ee
 We define $|\Mb|=\det \Mb$ as the determinant.
 Comparing with a multivariate Gaussian we see that  the matrix, $\Psib$, is the inverse of the data covariance matrix; 
   \be
  	\Psib = \Mb^{-1}.
 \ee
 As we shall find this matrix is central to our analysis, we shall define the inverse data covariance as the  {\it precision} matrix.
The model dependence on cosmological model parameters, $ \thetab$, may lie in either the mean, $\mub= \mub(\thetab)$, or the data covariance matrix, $\Mb=\Mb(\thetab)$, or both. 
   Throughout  we shall assume that the cosmological parameter dependence lies only in the mean.     
   In Appendix A we describe the data vectors commonly used in cosmological large-scale structure analysis: galaxy redshift surveys, cosmic microwave background experiments and weak lensing surveys. Throughout, we shall assume that the data is a set of power spectra estimated from the data, although of course our results hold for correlation functions and are general to Gaussian-distributed data.
   
%

 \section{Covariance and precision}
 
 \label{sec:cov_prec}
 

\subsection{ Data covariance matrix}

If we have a physical model for the covariance matrix, we would choose to use this. However, the statistical properties of the data may be poorly understood, for example the nonlinear regime for galaxy redshift surveys and weak lensing, and galaxy bias in redshift surveys,
or the data may have been processed in ways which are not straightforward to model analytically, e.g., 
in CMB data where long-wave variations in the time-ordered-data may have to be removed via polynomial fits, which can alter the statistical properties. In these cases we use an ensemble of simulations to estimate the sample data covariance matrix.  In surveys where we do not know how to accurately simulate the data, we can use the data itself to estimate the data covariance. We return to this issue in Section \ref{sec:8}.

 If we generate $N_S$ independent realisations of the data, $\D_\alpha$, where each realisation is labelled by a Greek index, $\alpha, \beta, \dots$,  and adopt a convention of labelling the data-vector so that the Roman indices, $i,j,\dots$, indicates the wavenumber, $\ell$, or wavevector, $\k$, and redshifts $z, z',\dots$, the data-vector averaged over the realisations is 
 \be
 	\overline{\D} = \frac{1}{N_S} \sum_\alpha^{N_S} \D_{\alpha}.
 \ee
 The expectation value of the data-vector  is
  \be
  	\lgl \D_\alpha \rgl= \left\lgl \overline{ \D} \right\rgl = \mub .
  \ee
For independent and identically distributed realisations, and where we can use a symmetry or binning of the data to average over $N_{\rm modes}$ with the same mean value, the accuracy of the estimate of the mean data-vector will scale as
  \be
  	\sigma (\overline{\D})  = \sqrt{\frac{1}{N_S N_{\rm modes}}}\,  \mub.
  \ee
 An unbiased estimator for the data covariance matrix is the sample data covariance, $\widehat{\Mb}$, from an ensemble of $N_S$ independent and identically distributed realisations;
  \be
  	\widehat{\Mb}= \frac{1}{\nu} \sum_\alpha^{N_S}
	 \Delta{\D}_{\alpha} \Delta{\D}^t_{\alpha}  ,
	 \label{eq:datcov}
  \ee
  where
  \be
  	\Delta{\D}_\alpha = \D_\alpha - \overline{\D}_\alpha
  \ee
  is the variation in the data for each realisation, and $\nu$ is the number of degrees-of-freedom in the ensemble. If the estimated mean of the data-vector is know to be the expected mean, $\overline{\D}_\alpha=\lgl \D\rgl$ then  
   \be
   	\nu=N_S.
\ee
However, if the mean is estimated from the data itself,  we reduce the number degrees-of-freedom by one, so that
 \be
 	\nu=N_S-1.
\ee
  
%

  \subsection{The Wishart distribution} 
   \label{sec:wish}
   
  The statistical properties of the sample data covariance matrix, assuming the variations in the measured field are Gaussian-distributed, are given by the Wishart distribution (Wishart, 1928), which generalises the $\chi^2$-distribution; 
  \be
	p(\widehat{\Mb}|\Mb,\nu,\eta) = 
	\left(\frac{  \nu^{\nu\eta/2 } | \Mb|^{-\nu/2} \,  | \widehat{\Mb}|^{\gamma/2}   }{2^{\nu \eta/2}  
	\Gamma_\eta[\nu/2]} \right) e^{- \small \frac{\nu}{2} \Tr \widehat{\Mb} \Mb^{-1}}  \!\!
	,
\ee
where $|\Mb| = \det \Mb$ is the determinant of $\Mb$, 
\be
	\Gamma_\eta\left(\frac{\nu}{2}\right) = \pi^{\eta(\eta-1)/4} \prod^\mu_{s=1} \Gamma\left[\frac{\nu}{2}
	+\frac{1-s}{2}\right]
\ee
is the multivariate Gamma function (see Appendix B1 for a definition),  $\eta=N_D$ is the size of the data-vector,  $\Mb$ and $\widehat{\Mb}$ are $\eta\times \eta$ matrices,  $\nu$ is again the number of degrees of freedom of $\Mb$, and  $\gamma=\nu-\eta-1$. We require that $\nu > \eta$, to ensure the estimated data covariance matrix is positive definite.

For a single data point, where $\eta=1$, the Wishart distribution is the reduced-$\chi^2$ distribution,
\be
	p(y|\nu) =  \left( \frac{\nu}{2}\right)^{\nu/2} \frac{ y^{\nu/2-1} }{\Gamma[\nu/2]}   e^{\small - \nu y/2}
\ee
where $y=\widehat{M}_{11}/M_{11}= \chi^2/\nu$, with mean $\lgl y \rgl = 1$ and variance $\sigma^2(y) = 2/\nu $.

The mean of the general Wishart distribution is
 \be
 	\lgl \widehat{\Mb} \rgl  = \Mb ,
 \ee
 showing it is indeed an unbiased estimate of the covariance matrix,
 while the covariance of $\widehat{\Mb}$ is
  \be
  	\lgl \Delta \widehat{M}_{ij} \Delta \widehat{M}_{mn} \rgl = 
	\frac{1}{\nu} (M_{im}M_{jn} + M_{in}M_{jm}).
  \ee
 This result can also be derived  from the Gaussian four-point function or directly from the Wishart distribution (see Appendix B2 where we calculate the characteristic function for the Wishart).

\subsection{The precision matrix and Inverse-Wishart}

  The simplest estimator for the precision matrix is
  \be
  		\widehat{\Psib} =  \nu \left[ \sum_\alpha^{N_S}
	 \Delta{\D}_{\alpha} \Delta{\D}^t_{\alpha} \right]^{-1} ,
	 \label{eq:prec_est}
  \ee
  where $\nu$ is the number of degrees-of-freedom. 
This estimator follows an inverse, or inverted, Wishart distribution (see e.g., Press, 1982),
\be
	p(\widehat{\Psib}|\Psib,\nu,\eta ) =
	\left(\frac{ \nu^{\nu\eta/2} |\widehat{\Psib}|^{-\beta/2} | \Psib|^{\nu/2} }{
	2^{\nu\eta /2}   \Gamma_\eta[\nu/2]  
	} 
	\right) 
	e^{-\frac{\nu}{2}{\small  \Tr \,\widehat{\Psib}^{-1} \Psib}}  \,\,,
\ee
where  $\beta=\nu+\eta+1$, and  $\eta=N_D$ is the size of the data-vector. We derive the Inverse-Wishart distribution in Appendix B3.

For a single data point, $\eta=1$, the Inverse-Wishart  reduces to the  inverse-$\chi^2$ distribution, 
 \be
 	p(x|\nu) = \left(\frac{\nu}{2}\right)^{\nu/2} 
			\frac{x^{-\nu/2-1}}{\Gamma[\nu/2]}  e^{-\nu/2x},
 \ee
where $x = M_{11}/\widehat{M}_{11} = \nu/\chi^2$.
 The  mean of this distribution is  
  \be
  	\lgl x \rgl = \frac{\nu}{\nu-2}, \hspace{1.cm}  \nu>2
  \ee
  and its variance is given by
  \be
  	\sigma^2(x) = \frac{2\nu^2}{(\nu -2)^2 (\nu-4)}, \hspace{1.cm}  \nu>4.
  \ee
  Immediately we see that the inverse distribution has different properties to $\chi^2$. Not only is the mean of the inverse-distribution biased high, $\lgl x \rgl >1$, but both the mean and variance can diverge. 
  
  We can understand the behaviour of the inverse-$\chi^2$ by considering the underlying Gaussian field. If $\Delta D_\alpha$ is the one data point, sampled $N_S$ times,  the sample variance is $\widehat{M}_{11} = \sum_\alpha \Delta D^2_\alpha/\nu$. As the $\Delta D_\alpha$ fields are Gaussian, they will fluctuate symmetrically around $\Delta D_\alpha=0$. Squaring and summing will produce positive values with mean $\lgl \Delta D^2_\alpha \rgl = M_{11}$. However the sample variance will scatter around this, bounded from below by zero. When we invert the sample variance, some of the values which are close to zero will become arbitrarily large. As there are no compensating small values, these large values will bias the mean of the precision matrix high, skewing the distribution.

  The expectation value of the sampled precision matrix is biased and, assuming the mean in unknown, given by (Kaufman, 1967; see Press, 1982; Anderson, 2003; see also Hartlap et al., 2007, for a first application to cosmology)
  \be
  	\lgl \widehat{\Psib} \rgl = \frac{N_S-1 }{ N_S - N_D - 2} \Psib.
	\label{eq:meanprecision}
  \ee
If $N_S > N_D+2$  is not satisfied then the values of $\widehat{\Mb}$ are not positive-definite and its inverse is undefined. If we do satisfy this condition  then the bias on the  inverse can be corrected to yield an unbiased estimate of the precision matrix given by
  \be
  	\Psib_{\rm unbiased} = \frac{ N_S - N_D - 2} {N_S-1 } \widehat{\Psib}.
	\label{eq:unbias_prec}
  \ee
 In fact, this estimator for the precision matrix is the only unbiased estimator. 
   
 The covariance of the sample precision matrix, again assuming the mean is unknown, is (Kaufman, 1967; see also Press, 1982, Matsumoto, 2011)
  \ba
 \lefteqn{\left\lgl \Delta \widehat{\Psi}_{ij} \Delta \widehat{\Psi}_{mn} \right\rgl =   }  \nn
		 & &
		 \hspace{-0.6cm}
		  A
		\Big[ 2  \Psi_{ij} \Psi_{mn} 
		+\,\, 
		(N_S-N_D-2) \left(\Psi_{im} \Psi_{jn} 
		+\Psi_{in} \Psi_{jm}   \right) 
		  \Big]    ,
 \label{invcovcov}
 \ea
  where
 \be
 	A= \frac{ (N_S-1)^2}{
		(N_S-N_D-1)(N_S-N_D-2)^2 (N_S-N_D-4)}.
 \ee
The second term in equation (\ref{invcovcov}) has the same form as the covariance for the data covariance matrix of Gaussian-distributed variables, and dominates when $N_S\gg N_D+2$.  This arises for large numbers of realisations as the Central-Limit Theorem will tend to make the Inverse-Wishart  Gaussian distributed.
The first term in equation (\ref{invcovcov}) arises from the shift in the biased  mean of the precision matrix.

 The covariance matrix of the sample precision matrix is also biased high. If uncorrected, it leads to an overestimate of the uncertainty in the precision matrix and parameter errors. 
We  can correct for the biases in the mean and covariance of the sample precision matrix, assuming the number of degrees-of-freedom is known and $N_S > N_D +2$. The unbiased covariance of the precision matrix can be found by substituting the prefactor, $A$, in equation (\ref{invcovcov}) by a corrected factor;  
   \be
 	A_{\rm corr}= \frac{ 1}{
		(N_S-N_D-1) (N_S-N_D-4)}.
 \ee
 The unbiased variance of the elements of the estimated precision matrix  is
   \be
  \sigma^2_{\rm corr} [\small{\widehat{\Psi}_{ij}} ] 
		\!=  \! A_{\rm corr} \big[ (N_S-N_D)  \Psi_{ij}^2 +  
		  (N_S-N_D-2)   \Psi_{ii}
		  \Psi_{jj}  \big] ,
 \label{errcov}
  \ee
  where no summation over repeated indices is implied. 
  A useful expression is the trace of this variance, which is given by
  \be
  	\Tr \, \sigma_{\rm corr}^2 [\widehat{\Psib}] = \frac{2}{(N_S-N_D-4)}  \sum_{i} \Psi_{ii}^2.
	\label{tracecov}
  \ee
  In the limit that $N_S \gg N_D+4$ the uncertainty in the precision matrix  falls off like the uncertainty on the data covariance matrix,  $\sqrt{2/N_S}$. However, unlike the sample covariance matrix,  the uncertainty on the sample precision matrix diverges when the number of realisations is close to the number of data points, while the sample covariance matrix is singular for fewer realisations. 
 Hence, we find two closely related requirements. The number of independent realisations used to estimate the data covariance and precision matrix, $N_S$, must be larger than the total number of data-points, $N_D$, being measured, $N_S > N_D +2$, to allow a correction for the bias in the estimated precision matrix, and $N_S > N_D +4$ to avoid a divergence in the error on the precision matrix.
Hence the bias and covariance of the precision matrix only depend on the number-of-degrees of freedom, $N_S - N_D$, and the model precision matrix, and are independent of the details of the experiment. This is a very simple, and powerful, result.


\section{An Example from Cosmology } 
\label{sec:example}

\subsection{Simulating a Weak Lensing Survey}

As an example of  the bias and variance of the sample precision matrix for a cosmological survey, we consider  a simulation of a weak lensing shear survey. In Appendix A3 we describe the weak lensing fields and power spectra. Here we consider a single shear field, with no $B(\beta)$-modes.
We generated  $N_S=100$ samples of a $10 \times 10$ square degree  weak lensing survey, with a Gaussian random shear field at a single redshift which we used to estimate the mean,  covariance and precision matrix of the shear power. The surface density of galaxies is $\bar{n}_2 = 30 $ per square degree, with a median redshift of $z_m = 0.9$.
We did this for sample sizes from $N_S=10$ to $N_S=100$. We repeated this 100 times to generate independent groups of the $N_S$ samples to estimate the mean and variance of the covariance and precision matrices.

This numerical experiment is non-trivial, as it has some more realistic assumptions compared to our analysis. In the simulations the underlying shear field is Gaussian-distributed, rather than the shear power itself. In practise this is what we expect for the large angular scale shear field, while on small scales we expect the shear field to be nonlinear and non-Gaussian. This will test if, on large scales, the assumption of Gaussian-distributed power is justified. 

The Gaussianity of the shear field for a single redshift ensures that the shear power covariance matrix is diagonal, $M_{ij}=M_{\ell \ell'} =M_\ell \delta^K_{\ell\ell'}$. The covariance matrix of the sample data covariance matrix is 
\be
\left\lgl \Delta \widehat{M}_{\ell} \Delta \widehat{M}_{\ell'} \right\rgl  = 
\frac{2}{N_S-1}  \, M_{\ell}^2  \, \delta^K_{\ell \ell'}  ,
\ee
while the covariance matrix of the unbiased precision matrix is 
  \be
 		\left\lgl \Delta \widehat{\Psi}_{\ell}\Delta \widehat{\Psi}_{\ell'}\right\rgl
		= 
		 2A_{\rm corr}
		\left[ 
		(N_S-N_D-2) \Psi_\ell^2  \, \delta^K_{\ell\ell'}
		 +   \Psi_{\ell} \Psi_{\ell'} \right]   .
 \label{invcovcovdiag}
 \ee
 The Gaussian part is diagonal, as expected, however the shifted term introduces off-diagonal terms. The diagonal of the precision matrix now propagates into every term in the precision covariance matrix, with the ratio of diagonal (Gaussian) to non-diagonal (shift-term) terms scaling as $N_S-N_D-1$.  The variance of the bias-corrected precision matrix is 
  \be
  	\sigma^2_{\rm corr}[\widehat{\Psi}_\ell] =  \left( \frac{2}{N_S-N_D-4}\right) \Psi_\ell ^2.
	\label{eq:varprec}
  \ee
 Taking the sum of this, or the trace of the covariance matrix, we recover equation (\ref{tracecov}).
 Gaussian-distributed  weak lensing convergence
  power spectra, on different angular scales and between redshift-bins, has the covariance  
  \ba
  	\lefteqn{\lgl |\Delta \widehat{C}^{\kappa\kappa}(\ell,z,z')|^2 \rgl = } \nn
	& &
	\hspace{-0.7cm}
	\frac{1}{f_{\rm sky}(2\ell +1)}\left[|C^{\kappa\kappa}(\ell,z,z')|^2 + C^{\kappa\kappa}(\ell,z,z) C^{\kappa\kappa}(\ell,z',z')\right]\!\! ,
  \ea
  where $f_{\rm sky}$ is the fraction of the sky covered by the survey. 
  If the shear power is binned into logarithmic passbands we divide by  $\ell \Delta \ln \ell = \Delta \ell$,  the number of $\ell$-modes in each passband.

\subsubsection{Whitening the covariance matrix}

 In cosmological surveys the data values can span several orders of magnitude, and so the conditional number of the corresponding covariance  matrix is very large. This can cause numerical instabilities in the inversion to estimate the precision matrix, as well as the failure of some non-standard estimators (see Section \ref{sec:opt_prec}). 
 If a good model of the covariance elements is known, one can whiten the covariance matrix (see e.g., Bond, et al., 1998) by rendering its diagonal elements close to unity. 
 Denoting the model data covariance elements by $M^{\rm mod}_{ij}$ and introducing a transformation matrix, $\calT_{ij}= (1/\sqrt{M^{\rm mod}_{ii}}) \delta^K_{ij} $, as the inverse of the square-root of the diagonal elements of the model covariance, we define a new, whitened, data covariance matrix by $\widehat{\Mb}_{\! W}= \calTb \widehat{\Mb} \calTb$. The whitened matrix, $\Mb_{\!W}$, has a conditional number close to unity and is readily inverted, which we carry out  using Singular Value Decomposition (SVD). The precision matrix is then obtained from the inverse of the whitened data covariance via $\Psib = \calTb^{-1} {\Mb}_{\! W}^{-1} \calTb^{-1}$. Here we whiten all data covariance matrices before correcting for whitening in the precision matrix.

\subsubsection{Numerical Results}

\begin{figure}
  \centering
  \includegraphics[width=1.0\columnwidth,clip=]{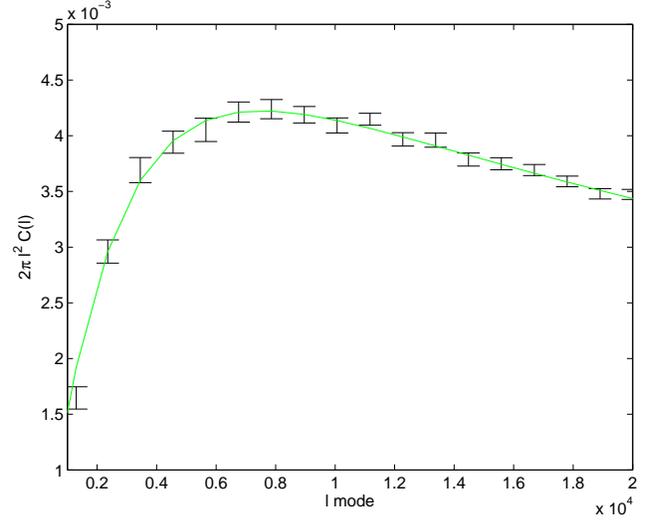}
  \caption{Simulated weak lensing shear auto-power spectrum, $C^{\kappa\kappa}(\ell,z,z)$ from 100 simulated $100$ square degree surveys, for a single median redshift of $z=0.9$. The solid line is the input power spectrum, while the data points are the mean estimated power spectrum, and the error bars are estimated from the 100 samples. We repeated this set of simulations 100 times to estimate the covariance of the data covariance and precision matrices.
    \label{fig:1}
 }
\end{figure}

Figure \ref{fig:1} shows the mean and scatter (diagonal of the covariance) in the estimated shear power spectrum from our $N_S=100$ simulated weak lensing survey, compared to the $\Lambda$CDM input model power.  We have chosen $N_D=36$ data points on the power spectrum as a compromise between oversampling the power spectrum with correlated data points, and under-sampling and missing some of its features which will contain parameter information.

\begin{figure}
  \centering
  \vspace{0.cm}
   \includegraphics[width=1.0\columnwidth,clip=]{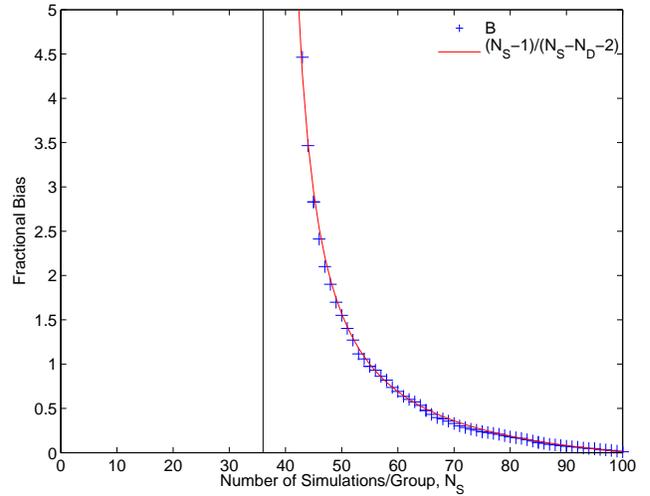}
	  \caption{The fractional  bias in the precision matrix of shear power spectra   from $N_S$ simulated and independent realisations of a $10^2$ square degree weak lensing survey, with $N_D=36$ power spectra data-points. The statistical properties of the precision matrix are generated from groups of 100 simulated surveys. The solid red line is the predicted scaling, while the vertical black line is the expected divergence for $N_S=N_D+2$. 
	  Blue crosses are the estimated bias from the simulations, using equation (\ref{eq:prec_est}), which closely follow the prediction.    We have suppressed error bars on points, which are correlated.
    \label{fig:2}
    }
\end{figure}

Figure \ref{fig:2} shows the fractional bias in the trace of the mean of the sample precision matrix, which we define as
 \be
 	{B} = \frac{\sum_\ell \lgl \widehat{\Psi}_\ell \rgl - \sum_\ell \Psi_\ell}{\sum_\ell \Psi_\ell} = \frac{N_S-1}{N_S-N_D-2},
 \ee
where $\widehat{\Psib}$ is estimated from equation (\ref{eq:prec_est}), for $N_D=36$ shear power spectra passbands as a function of number of realisations, $N_S$, in each group. The $N_S$ realisations are cumulative, so each point is correlated with points on the left.

 The numerical model, including the predicted divergence at $N_S=N_D+4$ (solid vertical line),  agrees extremely well with the prediction, as has previously been shown by Hartlap et al. (2007) in a similar cosmological context. There, they showed the bias followed the expected  behaviour using a simulated weak lensing survey based on ray-tracing through many lines-of-sight of the Millennium N-body simulation,  when the effects of non-linear clustering are included. The agreement between the simulations and prediction implies that the estimated precision matrix can indeed be debiased with the correction given by equation (\ref{eq:unbias_prec}).

\begin{figure}
  \centering
  \vspace{0.cm}
   \includegraphics[width=1.0\columnwidth,clip=]{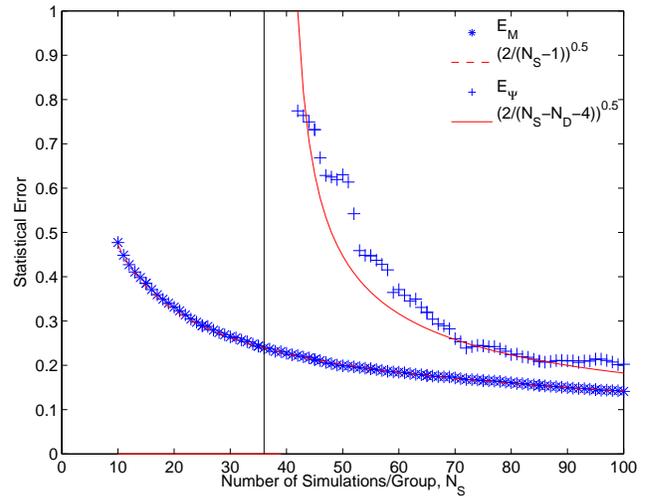}
  \caption{The error in the estimated precision and data covariance matrix from $N_S$ realisations of the Weak Lensing power spectrum, with $N_D=36$ data-points, generated from groups of 100 $10^2$ square degree simulated surveys, as a function of $N_S$.  The vertical black line is the number of data points. The blue stars are the statistical errors on the unbiased data covariance matrix, compared to the predicted scaling (dotted line). 
 Blue crosses are the statistical errors on the unbiased estimator of the precision matrix, equation (\ref{eq:unbias_prec}), compared to the predicted scaling. Again the agreement is good, with differences due to numerical effects. We have suppressed error bars on points, which are again correlated.
    \label{fig:3}
    }
\end{figure}

Figure \ref{fig:3} shows the measured fractional scaling of the trace of the covariance of the sample precision (crosses), defined as
 \be
 	{ E}_\Psi = \sqrt{\frac{\sum_\ell \sigma^2(\widehat{\Psi}_\ell)}{\sum_\ell \Psi_\ell^2}} = \sqrt{ \frac{2}{N_S-N_D-2}},
 \ee
 and the trace of the fractional error on the data covariance  matrix (stars), defined as
 \be
 	{ E}_M = \frac{\sum_\ell \sigma(\widehat{M}_\ell)}{\sum_\ell M_\ell} = \sqrt{\frac{2}{N_S-1}},
 \ee 
  as a function of number of realisations, $N_S$.  We have plotted the Wishart prediction (solid and dotted lines) for both statistics. The data covariance can be estimated for $N_S>1$ data points, and its variance is stable below the $N_S=N_D$ line. However, the variance of the precision matrix estimate diverges when we reach the number of data points, as predicted by the Inverse-Wishart distribution. Again the sample of $N_S$ realisations is cumulative and so each point is  correlated to the points on its left.  Again there is a very good agreement between the predicted and measured scaling of the variance of the data covariance matrix. The agreement between the predicted and numerical scaling of the variance of the precision matrix is also good, but there is some scatter and slight deviation which we attribute to the accuracy of the inversion of the sample data covariance.

  \subsection{The size of future surveys}
 
 As we have seen the main driver for the number of simulations comes from the precision matrix, whose accuracy is driven by the number of data points in our sample, $N_S  > N_D +4$. Here we discuss  typical values which will be encountered. The issue of data compression will be discussed in Section \ref{datacompress}.

\subsubsection{Pixelised or discrete data-sets}

In the case of pixelised data (Cosmic Microwave Background or Weak Lensing), or data where the individual data is sampled (Galaxy Redshift Surveys or Weak Lensing again) the number of data points can rise quite rapidly. If there is no analytic model for the pixel or data covariance matrix, the cost of simulations can be prohibitively high for $N_S > N_D$. We are then forced into some form of data compression, such as, in Cosmology, the estimation of two-point, or a number of n-point, power spectra or correlations. 

\subsubsection{Power spectrum analysis }

In the case of  a galaxy redshift survey, we imagine typically $N_k =50$ data points sampling the galaxy spectra. To sample the anisotropic distortion in redshift-space we would need $N_k^2$ data points. If this was repeated in $N_b=10$ redshift bins, we would have a total of $N_D = N_k^2 N_b = 2.5 \times 10^4$ data points in total.

  For a  tomographic, weak lensing power spectrum analysis measuring $N_{\rm spec}$ power spectra,  
  over $N_b$-redshift bins,  the total number of auto- and cross-spectra  for spin-2 fields (including $B$-modes) is 
     $
   	N_b (2 N_b +1 ).
   $
If  each power spectrum has $N_\ell$ passbands, the total number of data-points is 
 $	
 	N_D = N_\ell N_b (2 N_b +1 ) .
 $
If we again assume $N_b=10$, we can measure 210 different power spectra. With $N_\ell=50$ passbands per spectra per redshift,  we have $N_D=1.05 \times 10^4$ passbands, and the number of independent realisations we require is 
 \be
 	N_S > 1.05 \times 10^4 \left(\frac{N_\ell}{50}\right) \left(\frac{N_b}{10} \right)^2    .
 \ee
If we add the lensing magnification power, $C^{\mu\mu}(\ell,z,z')$, to this, the estimated number of independent realisations  increases by a significant factor. 

  If we combined cosmological probes we can estimate  the size of a combined data-vector. If we assume  $N_b=10$ redshift bins, we have a total of $N_f=43$ fields and
   \be
 	N_{\rm spec} = \frac{1}{2} N_f (N_f+1),
 \ee
auto- and cross-spectra. For our example  we then have $N_{\rm spec}=946$ spectra. Assuming further that each spectrum has $N_\ell = 50$ passbands, we have
  \be
  	N_D = N_\ell N_{\rm spec},
  \ee
 or  over  $5 \times 10^6$ data-points in our data-vector. These raw numbers clearly represent a significant challenge for generating realisations. 
In Section \ref{sec:8} we discuss ways in which to avoid the Wishart bound and the need  to generate such large numbers of simulated surveys.

 \section{Parameter covariance matrix} 
 
 \label{sec:paracovmat}

 \subsection{Covariance of the Fisher Matrix}

Having found the statistical properties of the sample precision  matrix we now turn to our main goal, to understand how the accuracy of the precision  matrix propagates 
 into  maximum likelihood parameter estimation. To do this, we use the Fisher matrix formalism (e.g., Tegmark, Taylor \& Heavens, 1997) to see how inaccuracies in the precision matrix  leads to inaccuracy in the Fisher matrix and leads to  inaccuracy in the parameter covariance matrix.

  The log-likelihood, $\calL \equiv-2 \log L$, can be expanded to second-order around its peak in parameter-space where the expectation value of the gradient of the log-likelihood is $\lgl \de_\alpha \calL \rgl =0 $, while the expectation value of the curvature yields the Fisher Matrix,
 \be
 			\lgl \de_\alpha \de_\beta \calL \rgl = 2 \calF_{\alpha\beta}.
 \ee
 The derivatives of the mean are taken with respect to the parameters. For Gaussian-distributed data, this is given by (Tegmark, Taylor \& Heavens, 1997)
   \be
 	\calF_{\alpha \beta} = \half  (\de_\alpha \mub \,\de_\beta \mub^t + \de_\alpha \mub^t \de_\beta \mub)
				\, \Psib.
 \ee 
 With the Gaussian approximation, the likelihood surface of the parameter-space is specified completely by the parameter covariance matrix, $\Phib$, given by the inverse of the Fisher matrix,
 \be
 	\Phi_{\alpha\beta} = \lgl \Delta \theta_\alpha \, \Delta \theta_\beta \rgl = \calF_{\alpha \beta}^{-1}.
 \ee
 If the data is again assumed Gaussian-distributed, the uncertainty on the precision  matrix propagates into the Fisher matrix by 
 \be
 	\Delta \calF_{\alpha\beta} = \half (\de_\alpha \mub \,  \de_\beta \mub^t+ 
		\de_\alpha \mub^t  \de_\beta \mub) \Delta \Psib,
 \ee 
 where $\Delta \Psib$ is a random variation in the precision matrix.  The covariance between terms in the Fisher matrices is given by
  \ba
  	\lgl \Delta \calF_{\alpha \beta}  \Delta \calF_{\mu \nu} \rgl \!\!\!\!  &=& \!\!\!\!
	\frac{1}{4} (\de_\alpha \mub \, \de_\beta \mub^t +\de_\alpha \mub^t  \de_\beta \mub )
	\lgl \Delta \Psib \Delta \Psib \rgl\, 
	\nn
	& &
	 \times \,\, (\de_\mu \mub^t  \de_\nu \mub+\de_\mu \mub \,  \de_\nu \mub^t).
  \ea
  We shall assume that the uncertainty in the mean of the data, $\mub$, is negligible. Substituting equation (\ref{invcovcov}) in for the covariance of $\Psib$ we find the unbiased covariance of the Fisher matrix is 
   \ba
 	\lefteqn{\lgl \Delta \calF_{\alpha\beta}\Delta \calF_{\mu\nu} \rgl 
	=  } \nn
	& & \hspace{-0.7cm}
    		A_{\rm corr}	\left[  (N_S-N_D-2)  
		\left(\calF_{\alpha\mu} \calF_{\beta \nu} 
		+  \calF_{\alpha\nu} \calF_{\beta \mu}  \right) +  
			2  \calF_{\alpha\beta} \calF_{\mu\nu}  \right],
		\label{fishcov}
 \ea
 valid for Gaussian-distributed data.

 \subsection{Covariance of the parameter covariance}

 The parameter covariance matrix is the inverse of the Fisher matrix  and so the uncertainty in the parameter covariance matrix is, to first-order,
 \be
	\Delta \Phi_{\alpha\beta} =
	- \calF^{-1}_{\alpha\gamma} \Delta \calF_{\gamma\delta}  \calF^{-1}_{\delta \beta} ,
 \ee
 where we assume summation over repeated indices. The covariance of the parameter covariance matrix is
 \be
 	\lgl \Delta \Phi_{\alpha\beta} \Delta \Phi_{\mu\nu} \rgl 	
	 =  
		 \Phi_{\alpha \delta} \Phi_{\eta \beta} 
		  \lgl \Delta \calF_{\delta \eta}  \Delta \calF_{\gamma \ep} \rgl
		 \Phi_{\mu \gamma} \Phi_{\ep \nu}  .
	\label{covpara}
 \ee
  Substituting equation (\ref{fishcov}) for the covariance of the Fisher matrix, we find
 the covariance of the parameter covariance matrix is 
 \ba
 \lefteqn{	\lgl \Delta \Phi_{\alpha\beta} \Delta \Phi_{\mu\nu} \rgl 
	 = A_{\rm corr} \big[  (N_S-N_D-2)  
		\left(\Phi_{\alpha\mu} \Phi_{\nu \beta}  
		+  \Phi_{\alpha\nu} \Phi_{\beta \mu}  \right)
		} \nn
		& & 
			\hspace{2.3cm}
		 +\,\, 2  \Phi_{\alpha\beta} \Phi_{\nu \mu}  \big].
 \ea
 This is a central result of this paper. 
 From this we see that the Inverse-Wishart covariance propagates through to the covariance of the parameter covariance matrix, with the Gaussian and shift terms.  Again this diverges if $N_S \le N_D +4$. 
The components of this matrix can be written as 
 \ba
 	\lgl |\Delta \Phi_{\alpha\alpha}|^2 \rgl 	
	& = &  
	 \frac{2}{N_S-N_D-4} |\Phi_{\alpha\alpha}|^2, 
	 \label{covcov1}
	\\
	 \lgl |\Delta \Phi_{\alpha\beta}|^2 \rgl	
	& = &  
	A_{\rm corr} \big[ (N_S-N_D) r_{\alpha\beta}^2  \nn
	& & 
	\hspace{1.0cm} + \,(N_S-N_D-2) \big] \Phi_{\alpha\alpha} \Phi_{\beta\beta}, 
	\\
	\lgl \Delta \Phi_{\alpha\alpha} \Delta \Phi_{\beta\beta} \rgl 	
	  & = & 
	2 A_{\rm corr} \left[1+(N_S-N_D-2) r_{\alpha\beta}^2 \right] \Phi_{\alpha\alpha}\Phi_{\beta\beta},	\nn
	\hfill
	\\
	\lgl \Delta \Phi_{\alpha\beta} \Delta \Phi_{\beta\beta} \rgl 	
	& = & 
	 \frac{2}{N_S-N_D-4}\Phi_{\alpha\beta} \Phi_{\beta\beta},
	 \label{covcov2}
 \ea
 where we have defined the parameter correlation coefficient,
 \be
 	r_{\alpha \beta} = \frac{\Phi_{\alpha\beta}}{\sqrt{\Phi_{\alpha \alpha} \Phi_{\beta\beta}}}.
 \ee
 From this result  we see that the main factors which affect the covariance of the parameter covariances are the difference between the number of realisations of the survey and the size of the dataset, $N_S-N_D$, the degrees-of-freedom of the precision matrix, and the parameter correlation coefficient, $r_{\alpha\beta}$.  This now provides us with a way to determine the accuracy with which we can estimate the distribution of parameter values in parameter space.

\subsection{Accuracy of parameter errors}

We can use this result to demonstrate how the accuracy of the errors on a parameter can be translated into the number of degrees-of-freedom in the data covariance, or equivalently the number independent realisations needed to estimate the precision matrix, and on the  accuracy on the precision and data covariance matrices. For a single cosmological parameter (marginalised over all other parameters), the error on the parameter variance is given by equation (\ref{covcov1}). Comparing this to equation (\ref{eq:varprec}), and  assuming that the data covariance is diagonal, we see that the fractional accuracy of the parameter variance is equal to the fractional accuracy of the precision matrix. 

\begin{figure}
  \centering
  \hspace{2.5cm}
   \includegraphics[width=0.5\columnwidth,clip=]{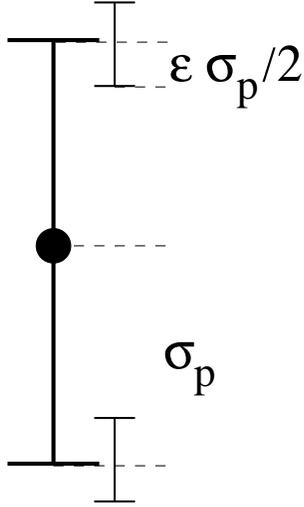}
  \caption{
  Sketch of the error on a parameter, $p$, given by $\sigma_p$, and the error on the error bar, $\epsilon \sigma_p/2$, where $\epsilon$ is the fractional variance on the precision matrix. 
    \label{fig:sketch_error}
    }
\end{figure}

 Defining $\varepsilon$ as the fractional accuracy of the parameter variance (equation \ref{covcov1}),
 \be
 	\varepsilon = \frac{\sigma[ \Phi_{\alpha\alpha}]}{|\Phi_{\alpha\alpha}|} = \sqrt{\frac{2}{N_S-N_D-4}} .
 \ee
  we can consider an arbitrary parameter, $p$, with expected value $p_0$. A measurement of $p$ will yield an uncertainty, $\sigma_p$, while the error on that uncertainty will be $\epsilon \sigma_p/2$. We can write this  as
  \be
  	p = p_0 \pm \sigma_p \left(1 \pm \half \epsilon \right).
  \ee
Figure \ref{fig:sketch_error} shows a sketch of this, illustrating for a single parameter the error bar and error on the parameter error.

\begin{figure}
  \centering
   \includegraphics[width=1.0\columnwidth,clip=]{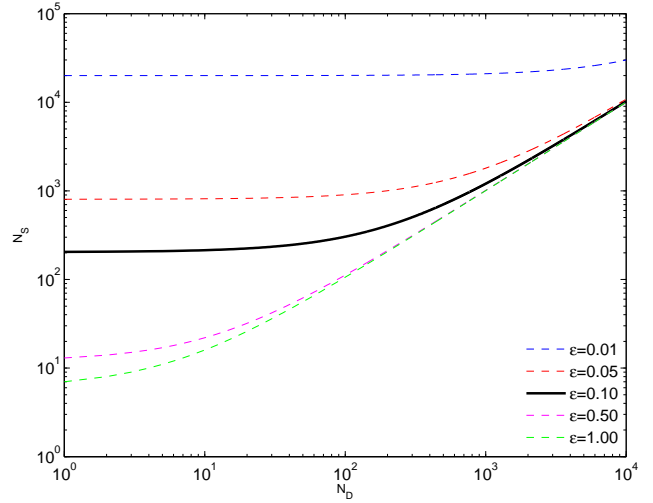}
  \caption{
  Scaling of the number of independent realisations of the survey, $N_S$, as a function of the size of the data set, $N_D$, for different fractional accuracies on the variance of the parameter variance, $\varepsilon$.
    \label{fig:NSvND}
    }
\end{figure}

 The  fractional error on the diagonal elements of the precision matrix is then given by (from equation \ref{errcov}, see also equation \ref{eq:varprec}),
  \be
  	 \frac{\sigma [\Psi_{ii}]}{|\Psi_{ii}|} = \varepsilon.
	 \label{eq:frac_acc_psi}
  \ee 
 Independent of the details of the survey,  we find  the required number of independent realisations of the survey for a given parameter accuracy and number of data points  is 
 \be
 	N_S >  \frac{2}{\varepsilon^2} +(N_D +4).
		\label{eq:Nsim}
 \ee
 The first term here {\bf is the} usual root-$N_S$ scaling for independent samples, and sets a lower limit on the number of independent realisation required to reach a given accuracy. The second term arises from the Inverse-Wishart variance, where the number of independent realisations needed to {\bf reach} a given accuracy scales as  the number of data points.
  Figure \ref{fig:NSvND} shows the scaling of the number of independent samples, $N_S$, with the size of the data-set, $N_D$. The value of $N_S$ in the limit $N_D \rightarrow 0$, is set by the desired accuracy of the parameter variance.

 The fractional error on the data covariance matrix, for a given parameter error accuracy and number of data points, is  
 \be
 	\frac{\sigma [ M_{ii}]}{|M_{ii}|}  < \sqrt{\frac{  2\varepsilon^2}{2 + \varepsilon^2 (N_D+4)}}.
 \ee
This has two regimes. When $\varepsilon^2 \ll 2/(N_D+4)$, i.e. for small data-sets compared to the required accuracy, the fractional error on the data covariance scales as 
 \be
 	\frac{\sigma [  M_{ii}]}{|M_{ii}|}  = \varepsilon,
\ee 
the same as for the precision matrix and the variance on the parameter variance, 
while for $\varepsilon^2 \gg 2/(N_D+4)$, when the data-set is large, the error on the data covariance scales as 
 \be
 \frac{\sigma[  M_{ii}]}{|M_{ii}|}   <  \sqrt{\frac{2}{N_D+4}} \ll \varepsilon.
 \ee
For large-data sets, this scaling puts the strongest constraints on the accuracy of the data covariance matrix, which can be much higher than the accuracies of the precision and parameter covariance matrices.

\subsection{Constraining the parameter error}
\label{sec:para_err}

While the accuracy of the parameter error  is set by the number of independent realisations of the survey, $N_S$, and the data size, $N_D$, it is useful to consider what typical accuracies any analysis should achieve, independent of the details of the particular survey. A reasonable accuracy for a parameter error is  $5\%$, since a much higher accuracy will put strong requirements {\bf on} the number of realisations, while lower accuracy will compromise the measurement error.  This requires that the marginalised parameter variance should be accurate to $10\%$ and that the precision matrix,  $\Psib,$ is accurate to $10\%$, or  $\varepsilon = 0.1$. From equation (\ref{eq:Nsim}), this requires $N_S > 200 + N_D +4$ independent realisations of the survey to reach this accuracy.  

 For a small data set, with $N_D \ll 100$,  a minimum of 204 independent realisations of the survey yields a $5\%$ error on the parameter error. This implies that the data covariance matrix is accurate to $5 \%$. When the data-set becomes $N_D \gg 100$, we require $N_S > N_D+4$ independent realisations, and the fractional accuracy of the data covariance matrix scales as 
  \be
  	\frac{\sigma[  M_{ii}]}{|M_{ii}|}   \approx  \sqrt{\frac{2}{N_D}} \ll 5\%.
  \ee
  In particular, for forthcoming tomographic cosmological surveys with 10 redshift bins, we can expect some $10^4$ power spectra data points requiring at least this number of independent realisations. This will increase the accuracy on the data covariance matrix to $1.4\%$. For combined data-sets the number of data points can rise to $N_D\approx10^6$, and hence require an accuracy of $0.14\%$ on the data covariance matrix. Achieving these accuracies will be challenging.

 
\section{Figures-of-Merit}

\label{sec:FoM}

\subsection{Uncertainty of the Figure-of-Merit}

In addition to marginalised parameter errors, it is useful to know how the uncertainty in the precision matrix affects the FoM. The FoM is the inverse of the enclosed area within a certain likelihood contour, and is frequently used as a target statistic to optimise cosmological surveys. Our aim here is to understand how inaccuracies in the precision matrix propagate through the parameter estimation into a FoM, and how fixing the required FoM can be used to put constraints on the accuracy of the precision matrix and data covariance matrix.

The dark energy Figure-of-Merit (DE FoM), $\Xi_{w_0 w_a}$, is defined as the inverse of the area of the 68\% error-ellipse for a two-parameter dark energy model (e.g., Albrecht et al., 2006),
 \be
 	\Xi_{w_0 w_a} = \frac{1}{\sqrt{\Phi_{w_0w_0} \Phi_{w_aw_a} - \Phi_{w_0 w_a}^2}},
 \ee
 where $w_0$ and $w_a$ parameterise the dark energy equation of state, $w(a) = \rho_{\rm de}(a)/P_{\rm de}(a)= w_0 + w_a (1-a)$ (Chevallier \& Polarski, 2001; Linder, 2003), where $a(t)$ is the cosmological sale factor, and  $\rho_{\rm de}$ and $P_{\rm de}$ are the energy-density and pressure of the dark energy. 
We  define a general FoM matrix for any two parameters as
 \be
 	 	\Xi_{\alpha \beta} = \frac{1}{\sqrt{\Phi_{\alpha \alpha} \Phi_{\beta\beta} - \Phi_{\alpha \beta}^2}}
		= \frac{1}{\sqrt{ (1-r_{\alpha\beta}^2)\Phi_{\alpha \alpha} \Phi_{\beta\beta}}},
 \ee
where no summation over the repeated indices $\alpha$ and $\beta$ is implied. In the second expression we have used the parameter correlation coefficient, $r_{\alpha\beta}$.

It is useful to consider the inverse of the elements of the FoM, the area of each ellipse in the parameter space
 \be
 	A_{\alpha\beta} = \frac{1}{\Xi_{\alpha\beta}}.
 \ee
The fractional change in $\Xi_{\alpha\beta}$ due to a change in the area is 
 \be
 	\frac{\Delta \Xi_{\alpha\beta}}{\Xi_{\alpha\beta}}  = - \frac{\Delta A_{\alpha\beta}}{A_{\alpha\beta}} .
 \ee
 Varying the parameter covariance matrix in the FoM, we find the fractional change in the area of the error ellipse is 
 \be
 	\frac{\Delta A_{\alpha\beta}}{A_{\alpha\beta}}  \!= \!\frac{1}{2(1-r_{\alpha\beta}^2)} \!
			\left(\! \frac{\Delta \Phi_{\alpha\alpha}}{\Phi_{\alpha\alpha}}\!+ \!
			\frac{\Delta \Phi_{\beta\beta}}{\Phi_{\beta\beta}} \!-\! 
			2r_{\alpha\beta} \! \frac{ \Delta \Phi_{\alpha\beta}}{\sqrt{\Phi_{\alpha\alpha}\Phi_{\beta\beta}}} 	
			\!		\right) \!\!.
 \ee
Using the results of the covariance of the parameter covariance matrix, equations (\ref{covcov1}) to (\ref{covcov2}) for the unbiased precision matrix, we find the variance of each FoM is
 \be
 	 \sigma^2 [\Xi_{\alpha\beta}]  =
	 \frac{(N_S-N_D)}{(N_S-N_D-4)(N_S-N_D-1)}  |\Xi_{\alpha\beta}|^2.
	 \label{eq:lnFom}
\ee
Once again, the fractional variance of the FoM depends only on the difference between the number of independent realisations of the survey used to estimate $\Mb$ and $\Psib$, and the size of the data set.

\begin{figure}
  \centering
  \vspace{0.cm}
  \includegraphics[width=1.0\columnwidth,clip=]{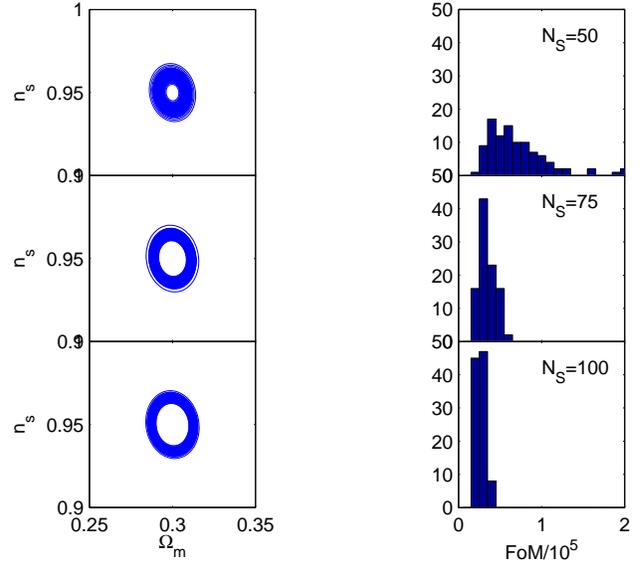}
  \caption{Numerical simulation of the parameter likelihood contours, estimated from a simulated weak lensing survey. The left-hand side (LHS) shows the 2-parameter 86.3\% likelihood  contour for the cosmological parameters, $\Omega_m$, the density parameter of matter and the clustering spectral index $n_s$. The right-hand-side (LHS) shows the frequency distribution of the FoM (inverse area) for these parameters over 100 realisations. The top row is for a sample-size of $N_S=50$ realisations, the middle row for $N_S=80$, and the bottom row for $N_S=100$. As the number of realisations increases the accuracy increases as predicted.  
   \label{fig:4}
}
\end{figure}

In Figure \ref{fig:4} (LHS)  we show the 2-parameter, marginalised  likelihood surface in the $\Omega_m - n_s$ plane for our weak lensing simulations. Each ellipse is a 2-parameter, $68.3\%$ likelihood contour for the group of simulations with $N_S$ realisations. As the number of realisations increases from $N_S=50$ to $100$, we see the spread in areas decreases. To quantify this, in Figure \ref{fig:4} (RHS) we plot the frequency distribution of the FoM for this parameter plane.

\begin{figure}
  \centering
  \vspace{0.cm}
  \includegraphics[width=1.0\columnwidth,clip=]{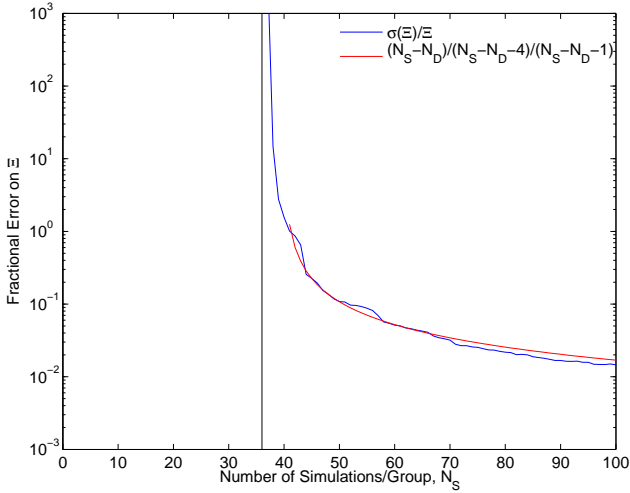}
  \caption{The scaling of the fractional error on the  FoM, $\Xi_{\alpha\beta}$, as a function of the  number of simulated realisations of a weak lensing survey, $N_S$, with $N_D=36$ data points. The solid blue line is the scaling predicted from the Inverse Wishart distribution, while the red line is the scaling found from the simulated surveys.   
   \label{fig:acc_fom}
}
\end{figure}

Figure \ref{fig:acc_fom} shows the predicted  uncertainty in the FoM,  equation (\ref{eq:lnFom}), compared to the variance of the  FoM distributions shown in the RHS column in Figure \ref{fig:4}, as a function of number of realisations, $N_S$. 
We see good agreement between our prediction and the error measured on the FoM from the weak lensing simulation.

\subsection{Accuracy of  the Figure-of-Merit}

In the design of many cosmological surveys, the FoM is used as a target statistic to optimise the survey design, varying area, depth and number of photometric passbands to find the design which maximises the FoM. Having set this optimal FoM, we then want to keep biases and uncertainties down to a level which does not violate the expected FoM. Here we develop an approach which uses the required FoM as a constraint to determine the number of survey realisations needed to do this. We then translate this into the accuracies of the precision and data covariance matrices. 

In order to keep within a required FoM we set the constraint that the uncertainty in the likelihood area, when added in quadrature with the area, should not exceed some fiducial value, $A^0_{\alpha\beta}$, 
 \be
 	A_{\alpha\beta}^2 + \sigma^2[ A_{\alpha\beta}] \le (A^0_{\alpha\beta})^2.
 \ee
Assuming the uncertainty in the area is small, and taking the expectation value, we can re-write this in terms of the FoM,
\be
	\Xi_{\alpha\beta} \left(  1  -
	\half  \left( \frac{\sigma[\Xi_{\alpha\beta}]}{|\Xi_{\alpha\beta}|} \right)^2  \right) 
	\ge \Xi^0_{\alpha\beta}.
\ee
The effect of a random change in the area of the error ellipse in parameter-space will, on average, reduce the FoM. Hence the actual  FoM we need  to measure, $\Xi_{\alpha\beta}$, to meet the required  $\Xi^0_{\alpha\beta}$ is increased. We define the fractional error in the FoM as
 \be
 	\varepsilon_{\small{ \Xi}}  \equiv   \frac{\sigma [ \Xi_{\alpha\beta}]}{|\Xi_{\alpha\beta}|}  .
 \ee
In order to keep the fractional increase in the FoM below some value, $\varepsilon_\Xi^2$, we can solve equation (\ref{eq:lnFom}) for $N_S$ and find that  the number of independent realisations should be  
 \be
 	N_S >  N_D + \frac{5}{2}     + \frac{1}{2 \varepsilon_\Xi^2}
		\left( 1 + \sqrt{(1+9 \varepsilon_\Xi^2)(1+ \varepsilon_\Xi^2)}\right).
		\label{eq:fom_NS}
 \ee
 In the limit that $\varepsilon_\Xi \ll 1$ we find,
  \be
  	N_S > N_D + \frac{1}{\varepsilon^2_\Xi}.
  \ee
If we want the fractional error on the FoM to be 10\%,  the number of realisation required (using equation \ref{eq:fom_NS}) is
 \be
 	N_S > N_D + 125.
 \ee
Again, we see that for small data-sets the number of realisations is fixed, this time at $N_S=125$, while for large-data-sets the number of realisations again scales as the number of data points.

  \subsection{Accuracy of the precision matrix}

 To set a constraint on the accuracy of the precision matrix, for a given accuracy on the FoM, we again only consider the diagonal components of the precision matrix, where $\sigma[\Psib_{ii}]=\varepsilon |\Psi_{ii}|$ (see equation \ref{eq:frac_acc_psi}). Substituting the constraint from the FoM on the number of realisations, equation (\ref{eq:fom_NS}), we find 
 \be
  \frac{\sigma^2[\Psi_{ii}]}{|\Psi_{ii}|^2} =  \frac{4 \varepsilon_\Xi^2}{1+ \sqrt{(1+9 \varepsilon_\Xi^2)(1+\varepsilon_\Xi^2)}-3 \varepsilon_\Xi^2},
  \ee
  which only depends on the ac curacy of the FoM.  In the high-accuracy regime, $\varepsilon_\Xi \ll 1$, this reduces to 
 \be
 	\frac{\sigma[\Psi_{ii}]}{|\Psi_{ii}|} \approx \sqrt{2} \varepsilon_\Xi.
 \ee  
 If we require for the FoM that $\varepsilon_\Xi =0.1$ this implies  that 
 \be
 	\sigma[ \Psi_{ii}]  \approx 0.19 |\Psi_{ii}|,
 \ee
or an accuracy of $19\%$ on the precision matrix.

\subsection{Accuracy of the data covariance matrix}

 Given we know that
 \be
  	 \frac{\sigma[ M_{ii}]}{|M_{ii}|} = \sqrt{\frac{2}{N_S}},
  \ee
  and that the number of independent realisations required to reach a given  accuracy of the FoM scales according to equation (\ref{eq:fom_NS}),  we can write 
 \be
 	 \frac{\sigma[ M_{ii}]}{|M_{ii}|} =       \frac{2 \varepsilon_\Xi}{\sqrt{1 +(5+2 N_D)\varepsilon_\Xi^2 + \sqrt{(1+9 \varepsilon_\Xi^2)		( 1+ \varepsilon_\Xi^2)} } }.
 \ee
 In the high-accuracy regime, $\varepsilon_\Xi^2\ll 1$, this reduces to 
  \be
  	 \frac{\sigma[ M_{ii}]}{|M_{ii}|}  \approx \sqrt{\frac{2 \varepsilon_\Xi^2}{1+N_D \varepsilon_\Xi^2}},
  \ee
  which for $N_D \varepsilon_\Xi^2 \ll 1$ reduces further to $\approx \sqrt{2} \varepsilon_\Xi$, scaling like the fractional accuracy of the precision matrix in the same regime, while  for $N_D \varepsilon_\Xi^2 \gg 1$ the fractional error reduces to $\sqrt{2/N_D}\ll\sqrt{ 2} \varepsilon_\Xi$. Hence, for high-accuracy FoM's and large data-sets, the accuracy of the data covariance matrix is driven by the size of the data-set. 
  
  For our fiducial accuracy of $\varepsilon = 0.1$ 
   we find 
  \be
  	\frac{\sigma[M_{ii}]}{|M_{ii}|} \approx \sqrt{\frac{0.02}{1+ (N_D/100)}}  ,
  \ee 
where for $N_D \ll 100$, the error on the data covariance is $\sigma[M_{ii}] \approx 0.19 |M_{ii}|$, the same accuracy as the precision matrix, while for $N_D \gg 100$  we find the $\sqrt{2/N_D}$ scaling.

\begin{figure*}
\centering
 \includegraphics[width=1.3\columnwidth,clip=,angle=-90]{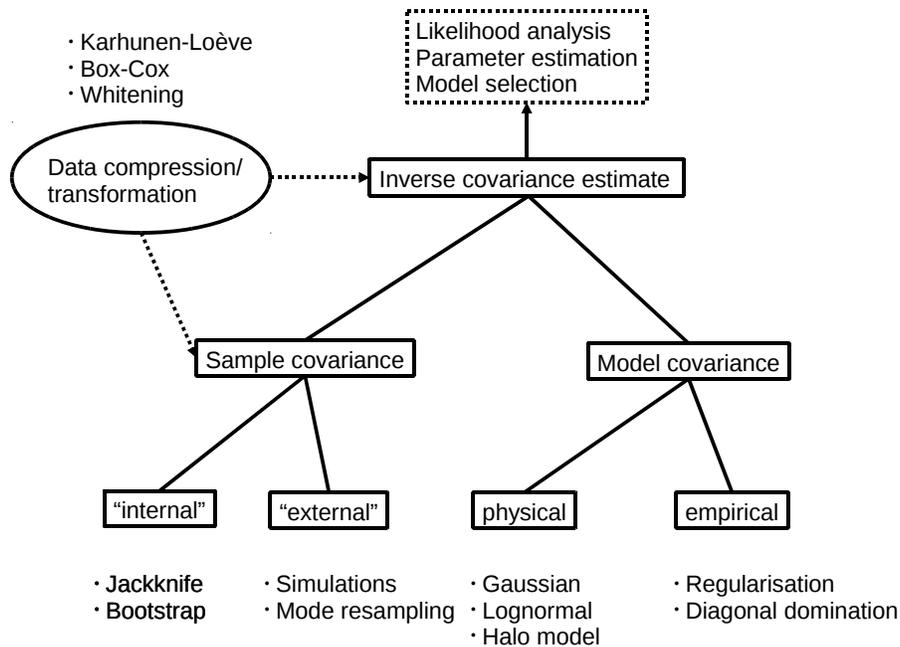}
 \hspace{-4.0cm}
 \vspace{-1.7cm}
\caption{
Different routes to determining inverse covariance matrices for use in likelihood analysis. The bottom level shows different ways to estimate sample and model covariance matrices, leading to the precision matrix and the likelihood function.}
\label{fig:sketch}
\end{figure*}

\section{Beyond the Wishart bound}
\label{sec:8}

The main conclusion of our analysis  is that without an accurate model data covariance matrix, and if we sample the data covariance, the number of independent realisations needs to be  greater than the number of data-points we are analysing. For large data-sets, such as  for the surveys now underway  in Cosmology, this requires  a prohibitively large number of simulations of the data.  
 In this Section we discuss  alternative routes to obtaining an accurate precision matrix which may help to meet, or avoid, the tight requirements set by simple estimation. Alternative approaches will also provide valuable consistency checks on some of our assumptions. We consider four methods:  theoretical modelling of the data covariance, optimal  estimators, data compression, and simulation and data resampling. Figure \ref{fig:sketch} shows  a schematic view of these methods and their relationship.    We shall not consider more radical alternatives, such as going directly to estimates of the likelihood function itself.

\subsection{Theoretical modelling of the data covariance}
\label{sec:theory_mod}

The problems of noise, inherent to the sample covariance matrix, are avoided altogether if we can accurately model the data covariance matrix  analytically. Modelling  of the data  covariance ranges from assuming the data is Gaussian distributed  on large-scales (e.g. Kaiser, 1992; Knox, 1995) to  assessing the impact of non-linear clustering,  using perturbation theory and simulations, on the galaxy power spectrum covariances (Meiksin \& White, 1999) and the weak lensing power spectra covariance (Scoccimarro, Zaldarriaga, Hui, 1999).  In addition, the halo model has been used to estimate the covariance matrix for the matter and weak lensing power spectra  (e.g., Cooray \& Hu, 2001; Takada \& Bridle, 2007; Takada \& Jain 2009;  Kayo et al., 2012), while estimation of  the data covariance matrix for a
lognormal  field (e.g., Hilbert, et al., 2011) seems to reproduce the main features of the covariance structure for weak lensing correlations.
Given the level of difficulty in modelling the nonlinear regime one needs to test the range of validity of these model against simulations. For example, 
Takahashi et al. (2009) have tested nonlinear modelling of the galaxy clustering data covariance matrix on a suite of 5000 simulations, while 
Sato et al. (2009) have tested nonlinear estimates of the weak lensing covariance on simulations. 
Kiessling et al. (2011) have also studied modelling non-Gaussian covariances and parameter forecasting using  weak lensing simulations. 
Hamilton \& Rimes (2005, 2006) first pointed out the loss of information in the quasi-nonlinear regime in the matter power spectrum due to non-Gaussianity using simulations, while Hamilton, Rimes \& Scoccimarro (2005) have discussed some of the issues with estimating the data covariance matrix from simulations. 
Theoretical models of the data covariance matrix have been applied to parameter estimation from data with the CMB (e.g. Verde, et  al., 2003; Spergel, et al., 2003), galaxy redshift surveys (e.g. Ballinger, et al., 1995; Tadros, et al., 1999), and weak lensing (e.g., Brown, et al., 2003; Kitching, et al., 2007).

Even if modelling  is not precise, one could develop analytic parameterised fitting functions to the data covariance, fitting the free parameters to simulations or the data. Importantly,  these physically motivated models could allow one to incorporate the cosmology dependence of the covariance, which would require a substantial increase in the number of simulated realisations to cover parameter space. Sometimes, truncation or smoothing of the sample data covariance is used to suppress noise (e.g. Mandelbaum et al., 2012). However, such approaches alter the number of degree-of-freedom in the data covariance and so we would no longer know how to correct the precision matrix for bias. 

If we assume for the moment that we can model the theoretical uncertainty on the data covariance matrix as random, even for the next generation of surveys when we expect $N_D \approx 10^4$, the data covariance has to be known to a few precent accuracy, which will become a problem for theoretical computation. For surveys with $10^6$ data points the accuracy of the data covariance has dropped to a fraction of a percent, putting high demands on its calculation (see Section \ref{sec:para_err}).

\subsection{Optimal precision estimators}
\label{sec:opt_prec}

If we do not have a reliable model or fitting function to the data covariance, we can still suppress the noise in the sample covariance by combining it with some simple model or prior knowledge. This is generally referred to as shrinkage estimation.  One can define optimised covariance estimators in the sense that they yield smaller variance than equation (\ref{errcov}) while keeping the bias small. We investigate the performance of three well-known cases.

\subsubsection{Shrinkage: Stein precision estimator}

Stein et al. (1972) have proposed the precision estimator
\be
	\widehat{\Psib}_{\rm Stein} = \frac{N_S -N_D -2}{N_S - 1}\; \widehat{\Psib}
		+ \frac{N_D(N_D+1)-2}{(N_S-1)\; {\rm Tr}\, \widehat{\Mb}}\; \mathbf{I}\;,
	\label{eq:stein}
\ee
which is defined for $N_S > N_D+2$.  If $N_S \gg N_D$, the estimator reduces to the unbiased estimate, $\widehat{\Psib}_{\rm unbiased}$. If $N_S \sim N_D \gg 1$, the estimator returns $S^{-1}\, \mathbf{I}$, where 
 \be
 	S= \frac{1}{N_D} \Tr \, \widehat{\Mb}
\ee
 is the average of the diagonals of $\widehat{\Mb}$. This is exact if the covariance is diagonal and homoscedastic. This estimator has smaller loss than any estimator that is proportional to $\widehat{\Psib}$ for a \lq natural\rq\ loss function (Stein, et al.,  1972), a generalisation of least squares between the matrix elements of the estimator and the true precision matrix.

\subsubsection{Shrinkage: Haff precision estimator}

A second estimator, suggested  by Haff (1974), is;
\be
	\widehat{\Psib}_{\rm Haff} = \frac{N_S -N_D -2}{N_S - 1} \left( (1 - \sqrt{U}) \widehat{\Psib}
	\!+ \sqrt{U}\; S^{-1}\; \mathbf{I}\right) ,
\ee
with
\be
	U = S^{-1}|\widehat{\Mb}|^{1/N_D}
\ee
again defined for $N_S > N_D+2$ only. The variable $U$ measures disparity among the eigenvalues of $\widehat{\Mb}$ and lies in the interval $0 \leq U \leq 1$, shifting from the unbiased sample estimator ($U=0$) to the estimator $\propto S^{-1}\, \mathbf{I}$ ($U=1$). This estimator has smaller loss than any estimator that is proportional to $\widehat{\Psib}$ for a whole class of loss functions (Haff, 1974). However, this property is only guaranteed close to the divergent case, in our case for $N_S \leq N_D + 4$.

\subsubsection{Target data covariance shrinkage}
\label{sec:target}

Shrinkage in its narrower sense refers to covariance estimates in which the balance between the sample covariance and the assumed model (the \lq target\rq, see Section \ref{sec:theory_mod}) is estimated from the data as well. The estimate for the precision matrix is given by the inverse of
\be
	\widehat{\Mb}_{\rm shrink} = \lambda\; \Tb + (1-\lambda)\; \widehat{\Mb}\;,
\ee
where $\Tb$ is the theoretical target covariance matrix. This formalism has been applied to covariance estimation of galaxy clustering power spectra by Pope \& Szapudi (2008), and to the CMB by Hamimeche \& Lewis (2009).  Ledoit \& Wolf (2003) derived an analytic estimator for the shrinkage intensity $\lambda$, thereby greatly reducing the computational cost of this form of shrinkage estimation. It is given by (see also Sch{\"a}fer \& Strimmer, 2005)
\be
 	\label{eq:lambdashrink}
	\lambda = \frac{\sum_{ij}^{N_D} {\rm Var}[ \widehat{M}_{ij} ] - {\rm Cov} [ T_{ij}, \widehat{M}_{ij} ]}
	{\sum_{ij}^{N_D} {\rm Var}[ \widehat{M}_{ij} - T_{ij} ] + \big( \overline{\widehat{M}}_{ij} - \overline{T}_{ij} \big)^2  }\;,
\ee
where the variances and covariances are computed from the $N_S$ realisations, so then (see also Pope \& Szapudi, 2008)
 \be
 	{\rm Var} [ \widehat{M}_{ij} ] = \frac{N_S^2}{(N_S-1)^3} \sum_{\alpha=1}^{N_S} \left( W_{ij}^{(\alpha)} -  		\overline{W}_{ij} \right)^2\;,
 \ee
where
 \be
	W_{ij}^{(\alpha)} = \Delta {D}_{\alpha,i}\; \Delta {D}_{\alpha,j}\;.
 \ee
As only one set of realisations is available in practice, the means in the denominator of Equation (\ref{eq:lambdashrink}) have to be replaced with estimates of $\widehat{\Mb}$ and $\Tb$. If the target matrix is noise-free, ${\rm Cov}[ T_{ij}, \widehat{M}_{ij} ]=0$ and ${\rm Var}[ \widehat{M}_{ij} - T_{ij} ]={\rm Var}[ \widehat{M}_{ij} ]$. If $\overline{\widehat{M}}_{ij} - \overline{T}_{ij}=0$,  the target accurately describes the covariance in the data and $\lambda=1$. Conversely, the shrinkage intensity tends to zero if the target attains a similar noise level as, and/or if the mean target deviates strongly from the mean of, the sample data covariance. 

We consider two choices for our target matrix to test on our weak lensing simulations. The first is a theoretical estimate of the covariance matrix based on a Gaussian-distributed power spectrum (e.g., Kaiser, 1992),
\be
	\Tb_1 = \left(\frac{C_i^2}{f_{\rm sky}\, \ell_i^2\, \Delta \ln \ell}\right) \I\;,
\ee
where $f_{\rm sky}$ is the fraction of the sky covered by the survey, and we have assumed log-binning. Since this should correspond closely to the simulations we have generated we expect $\lambda \rightarrow 1$. The second is an empirical target matrix, 
\be
	\Tb_2 = S  \I \;,
\ee
 which represents a minimum-knowledge approach.

\subsubsection{Testing Shrinkage}
\label{sec:shrinktest}

\begin{figure}
  \centering
  \vspace{0.cm}
   \includegraphics[width=1.0\columnwidth,clip=]{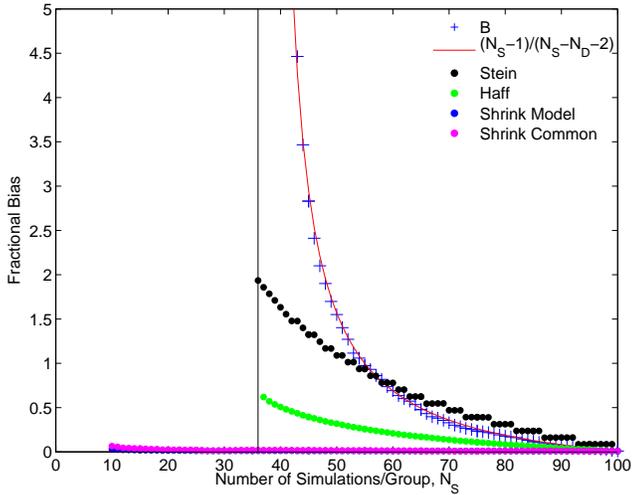}
  \caption{The bias in the estimated precision matrix from $N_S$ realisations of the Weak Lensing power spectrum, with $N_D=36$ data-points, generated from groups of 100 $10^2$ square degree simulated surveys, as a function of $N_S$.  The blue circles are for direct inversion of the unbiased data covariance matrix, while green are for the Stein and Haff estimators.
    \label{fig:bias_op}
    }
\end{figure}

\begin{figure}
  \centering
  \vspace{0.cm}
   \includegraphics[width=1.0\columnwidth,clip=]{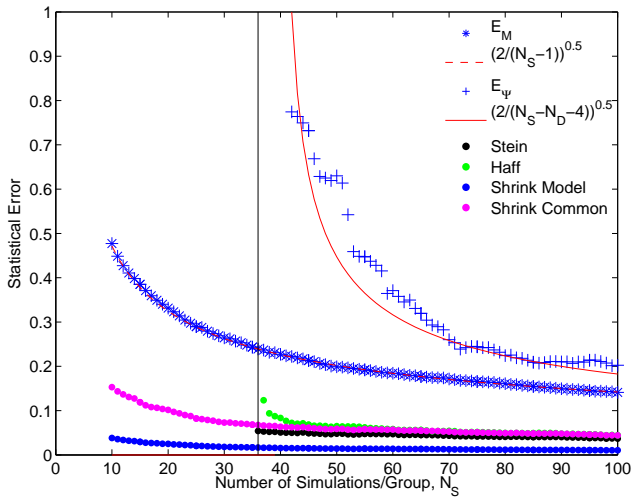}
  \caption{The error in the estimated precision matrix from $N_S$ realisations of the Weak Lensing power spectrum, with $N_D=36$ data-points, generated from groups of 100 $10^2$ square degree simulated surveys, as a function of $N_S$.  The blue circles are for direct inversion of the unbiased data covariance matrix, while green are for the Stein and Haff estimators.
    \label{fig:var_op}
    }
\end{figure}

 Figures \ref{fig:bias_op}  and \ref{fig:var_op} show the fractional bias and error on the precision matrix for the Stein estimator (black points), the Haff estimator (green points), and the two target shrinkage methods, model (blue points) and mean (purple points), applied to our weak lensing simulations . 
  
  
  \begin{itemize}
 \item {\bf Stein estimator:}
The Stein estimator is more biased than the simplest sample estimator for large numbers of realisation, even though it asymptotes to become the same estimator. Simulations would be required to calibrate this for any particular experiment.  The variance of the Stein estimator is low, yielding an almost  constant $5\%$ error.
 
\item{\bf  Haff estimator:}
The Haff estimator is less biased than the Stein, and the sample estimator, but still shows significant bias. The variance of the Haff estimated precision matrix is again around $5\%$. 

\item{\bf  Target estimator:}
The $\Tb_1$ model target shrinkage estimator is essentially unbiased, as we expect for an accurate model, and works for $N_S < N_D+4$.   The $\Tb_1$ model also does best at lowering the statistical to around $1\%$ accuracy.
The $\Tb_2$, empirical estimator yields a similarly unbiased precision matrix, which is of interest. The error on the estimated precision matrix is slightly higher than the model target, at $5\%$, similar to that for the Stein and Haff estimators. 
\end{itemize}

We conclude that the model  estimator, $\Tb_1$, works impressively well when the model is a good approximation,  minimising the bias in the precision matrix and reducing the error in the precision matrix to a few percent. For more realistic simulations, we would expect to take advantage of the more detailed theoretical models (see Section \ref{sec:theory_mod}).
 Interestingly, the  empirical $\Tb_2$ target is similarly unbiased, with a $5\%$ error, although we caution that our simulation covariance is also diagonal. Both of these estimators would satisfy our goal of an unbiased parameter errors with $5\%$ error, requiring only $N_S \approx 30$ realisations compared to the $N_S > 240$ needed for the sample data covariance. We imagine this would work just as well for much larger data-sets. 
The Haff and Stein estimator have a bias similar to the sample estimator, but without the known correction factor. One would have to calibrate these with simulations. If this can be done, the error is sufficiently low to be useable.

\subsection{Data compression}
\label{datacompress}

Independently of how the realisations to compute the sample covariance matrix are obtained, one can lower the Wishart bound by reducing the size of the data vector. Karhunen-Lo{\`e}ve eigenvalue methods (Tegmark, Taylor \& Heavens, 1997, and references therein) are widely used in astronomy to compress data by finding a smaller data vector that maximises the Fisher information (preserving parameter information), while simultaneously  diagonalising the new data-vector covariance. Heavens, et al. (2000) introduced a linear implementation that is lossless for an arbitrary number of estimated parameters if the data covariance is not parameter-dependent, and otherwise still performs better than principal component analysis of the Fisher matrix. Data compression can also help stabilise the inversion of the data covariance, since the noisy modes which cause numerical instabilities are removed (e.g., Taylor, et al., 2001).

We can consider what optimal compression may achieve  based on general considerations.
 The compression method of Heavens et al. (2000) compresses data to one  element per non-degenerate cosmological parameter. Future surveys will constrain cosmologies with a large number of parameters together with a substantial list of calibration and nuisance parameters, so that one can expect several tens of parameters in total, corresponding to a compression to $N_D \sim 100$.
Alternatively, one can argue from the number of characteristic features in the matter power spectrum, which features a number of transition scales beyond the overall amplitude and slope, and  the clustering growth rate, leading to an estimate of several tens of parameters.

\subsubsection{Compression in CMB analysis}
 Gupta \& Heavens (2002) demonstrate that the some 400 temperature power spectrum modes can be compressed to the $10-20$ cosmological parameters of interest. The computation of these modes requires a one-off $O(N_{\rm para} N_D^3)$ calculation, compared to the $O(N_D^3)$ needed for a brute-force likelihood approach. However our aim here is to minimise the uncertainty on the likelihood function, rather than to speed up the analysis.

\subsubsection{Compression in weak lensing}

 In lensing the COSEBI two-point statistics are an efficient way of compressing angular weak lensing data. Asgari et al. (2012) found that of the order 10 modes capture the bulk of the cosmological information. Moreover, the B-mode is expected to have low signal-to-noise while cross-correlations between E- and B-mode should vanish altogether, so that these signals can be compressed efficiently into a small number of elements.

For radial lensing modes, Heavens, et al. (2003) applied Karhunen-Lo{\`e}ve methods to a 3-D weak lensing analysis and found that only 4 (out of 100) radial modes contain significant information (see Hu, 1999, for similar conclusions on tomographic weak lensing data). Tomographic or 3-D weak lensing analyses will also have to simultaneously model intrinsic galaxy alignments, which are primarily separated from the lensing signal via their different redshift dependence. A maximum number of 10 radial elements in the data vector per angular frequency should be a conservative estimate. The main issue is then the independence of these modes, since we have of the order $\sim N_b^2$ cross-spectra. If the modes are independent, we only need around $N_D \sim 100$ spectra to consider, in agreement with the estimate for Karhunen-Lo{\`e}ve methods. If they are not truly independent and we rely on angular compression, we may only compress weak lensing data by a factor of a few. This gives us a range of possible compression  factor from  $10$ to a few. 

If data compression can compress data so that $N_D \approx N_{\rm para} \approx 100$, this would be very powerful and imply we only ever need around $N_S \approx 300$ realisations for any survey. However, this is probably over-optimistic, and if the real compression is a factor of 10, large data-sets of $N_D \approx 10^4 - 10^6$ will still be difficult to accurately analyse.

\subsection{Resampling techniques}

In previous Sections we have considered how we can reduce the need to generate large numbers of realisations of our surveys to ensure the accuracy of parameter errors. Here we discuss how we could generate these realisations. In general we can divide this into external realisations, usually from simulations, or internal realisations from e.g., Jackknife or Bootstrap resampling of the data itself.

\subsubsection{Simulation Mode Resampling}

If we do need to create large numbers of external samples via cosmological simulations, Schneider et al. (2011) have  proposed a method to rapidly generate, pseudo-independent random  realisations from a single N-body simulation.
 This resampling the large-scale, quasi-Gaussian Fourier modes using a semi-analytic formalism. The approach reproduces power spectrum covariances well, including the coupling between linear and non-linear scales, although a small bias in the covariance elements is introduced. Schneider et al. (2011) find that the number of full N-body simulations required to achieve the same error tolerance on the covariance matrix is reduced by a factor of 8, at the price of having to run these simulations into the future and with more frequent snapshot outputs.

 \subsubsection{Internal resampling: Jackknife}

If the statistical properties of the data are poorly known, one can create internal samples by resampling the observed data itself. In this case the data covariance is only estimated at one point in parameter-space, from a single realisation. 
A long-established method is the Jackknife method (Tukey, 1958), which, in the astronomical context of a correlated spatial random process, requires the survey to be split up into $N_{\rm sub}$ equally sized sub-regions. Jackknife samples are constructed by deleting one sub-region in turn (the delete-one Jackknife) and using the galaxy catalogues of the remaining survey area to re-compute the signal mean. The Jackknife covariance of this mean is then given by (Efron, 1980)
\be
\label{eq:jkcov}
	\widehat{\Mb}_{\rm Jack} = \frac{N_{\rm sub}-1}{N_{\rm sub}}\; \sum_{\alpha=1}^{N_{\rm sub}}
	 \Delta {\D}_{\alpha}\; \Delta {\D}_{\alpha}\;,
\ee
with $\Delta {\D}_{\alpha} = {\D}_{\alpha} - \overline{{\D}}_{\alpha}$, where the subscript $\alpha$ indicates both the realisation and  that the sub-region $\alpha$  has been deleted.
In the limit of uncorrelated data equation (\ref{eq:jkcov}) is equivalent to the standard estimator, equation (\ref{eq:datcov}) applied to the survey sub-regions (Efron, 1980; Shao \& Wu, 1989). In this case  there is no advantage in using Jackknifing. If there are correlations between sub-regions, these will be missed by standard estimation while the Jackknife, taken over the whole survey bar one sub-region, will measure these. 

We can estimate the number of sub-regions required for the Jackknife, by using the results for the Wishart distribution, assuming that correlations between the sub-regions are negligible, where $N_{\rm sub}=N_S$. For example in the case of a  weak lensing survey covering $15,000\,{\rm deg}^2$, the requirement of having of order $N_S=10^4$ realisation implies that the sub-regions would be little more than $1\,{\rm degree}$ on a side\footnote{Note that we do not consider to split up the survey into sub-\textit{volumes} as done in Norberg et al., (2009) because of the very strong correlations expected along the line-of-sight for a weak lensing survey, due to photometric redshift errors and particularly the broad lensing kernel.}. To avoid bias due to the impact of sub-region boundaries, the largest scales that could be probed would have to be much smaller, and hence jackknife estimates are likely restricted to, but potentially useful on, smaller scales. Even in this limit, the estimator based on equation (\ref{eq:jkcov}) will still be biased because the sub-regions are not independent, due to residual correlations, so that the actual number of degrees of freedom is smaller but unknown. In this case, we do not know how to correct for the Wishart bias.

 \subsubsection{Internal resampling: Bootstrap}

Another class of resampling techniques, containing a large number of variants, is the Bootstrap (Efron, 1979). This assumes that the empirical distribution function of a sample of independently and identically distributed data of size $n$ provides an unbiased estimate of the true underlying population from which the data has been drawn. Point estimates, e.g. of the covariance, can be obtained via standard estimators such as equation (\ref{eq:datcov}), by generating new samples from this empirical distribution, where the sum now runs over the number of bootstrap samples, $N_B$. If the bootstrapped sample has size $N_r$, the maximum number of distinct samples that can be drawn is $(n+N_r-1)!/[(n-1)! N_r!]$, which quickly grows large for moderate $n$ and $N_r$, so that usually Monte Carlo methods are employed to create bootstrap samples.
Sub-regions of the survey are resampled, analogous to the jackknife. In astronomy these resampled sub-regions are generally chosen to be non-overlapping (fixed-block bootstrap), although this may be suboptimal (Nordman, et al., 2007). 

If $N_B$ is large, the uncertainty on the covariance estimate becomes negligible, so that its inverse is an unbiased estimator of the precision matrix. However, there are multiple other sources of bias in the bootstrap technique which are hard to quantify. If the number of sub-regions is small, the empirical distribution is a coarse representation of the underlying population, and any  local features may be missed, and the convergence of the error tolerance is slower (Nordman, et al., 2007). If  the number of sub-regions is large and their area small, the correlation structure is not well preserved. It is possible that some sub-regions are not drawn at all in a given bootstrap sample, so that the total area coverage of this sample is less than that of the original survey, biasing the covariance estimate upwards. Drawing more sub-regions for bootstrap samples than the $N_{\rm sub}$ regions of the original data, i.e. $N_r > N_{\rm sub}$, can remedy this bias (Norberg, et al., 2009), but the choice of $N_r$ constitutes another parameter that requires calibration.

While the bootstrap method effectively by-passes the Wishart bound by creating a very large number of realisations, its multiple sources of bias, which all depend on the data set at hand, make an application to precision measurements questionable. Note that this is an active field of research, so that this conclusion could change on moderate time scales (see e.g. Loh, 2008, for a recent application of a spatial bootstrap variant to large-scale structure data).

\begin{table*}
  \begin{center}
ÊÊÊÊ\begin{tabular}{c l c c c }
	\hline
	ÊÊÊÊÊÊÊÊ	 {\bf METHOD}		& 			& 	{\bf Gain} 	& {\bf Cost} 	& 	{\bf Comment} \\ \hline
ÊÊÊÊÊÊÊ	{\bf Unbiased} 	&			& Unbiased, known variance  &	$N_S> 200+ N_D+4$&	Costly for large $N_D$	\\ \hline
	{\bf Modelling} 	&			& Unbiased/No variance  &	Complex modelling &	Nonlinear, baryonic physics	\\ \hline
Ê					&Ê{\bf Stein }		&factor $\approx 5$ reduction in $N_S$& 	Unknown bias 	& 	Range $N_S>N_D $ 	\\ 
	{\bf Shrinkage} 	& {\bf Haff 	}	& factor $\approx 5$ reduction in $N_S$	& 	Unknown bias 	& 	Range $N_S>N_D$	\\ \cline{2-5}
ÊÊÊÊÊÊÊÊ				&{\bf Model Target} 	& factor $\approx 10$ reduction in $N_S$	& 	Low/no bias  	& 	Applicable  for $N_S<N_D$	\\
ÊÊÊÊÊÊÊÊ				&{\bf Mean Target }	& 	factor $\approx 5$ reduction in $N_S$	& Low/no bias & Applicable  for $N_S<N_D$	\\  \hline
ÊÊÊÊÊÊÊ	{\bf Compression} 	& 			& 	$N_D \approx N_{\rm Para}\sim 100 $ 		& 	Information loss  	& Model dependent	\\ \hline
ÊÊÊÊÊÊÊÊ				& {\bf Simulation} 	& 	Factor $\approx 8$ gain 		&	More expensive simulations  	& 	Accuracy to be tested	\\ \cline{2-5}
ÊÊÊÊÊÊÊÊ	{\bf Resampling }	& {\bf Jackknife} 	&	 No simulations required 		& 	Unknown bias and variance  	& 	Needs calibration	\\
ÊÊÊÊÊÊÊÊ				&{\bf Bootstrap }	&	No simulations required /No bias	& Multiple unknown biases  	& 	Needs calibration 	\\
	\hline
ÊÊÊÊ\end{tabular}
\end{center}
 \caption{Summary Table of results of different methods for estimating the precision matrix for parameter estimation. The methods are described in detail in Section \ref{sec:opt_prec}, and the results of the tests in Section \ref{sec:shrinktest}.	
 }
 \label{surveytable}
\end{table*}

\subsection{Summary of Alternatives}

Which of the various routes to inverse covariance estimation is optimal depends strongly on the problem at hand, being influenced by aspects as diverse as the survey characteristics, the complexity of obtaining the signal, the computational cost of acquiring simulated survey realisations, the availability and accuracy of models, and the questions one aims to answer with the data. Clearly, if one has an accurate model of the data covariance to hand, this is what should be used. A poorer model can still be used, either to aid a fitting-function approach, or as the model target for shrinkage. Even if a model is not available, empirical shrinkage appears to work well, and all of these methods seem unbiased and yield a statistical error on the precision matrix less the 5\% even when $N_S < N_D$. Stein and Haff shrinkage  yield a bias which must be calibrated. Data compression may optimally reduct the number of data-points to the number of model parameters, typically a few hundred values, at the expense of loss of some information, and model-dependency. However, if not optimal the gain may only be a factor of 10's. Mode-resampling of simulations may lead to production of large numbers of realisation, with a factor of $\approx 8 $ possible. Empirical resampling of the data will lead to bias estimates which will need calibration, but the Jackknife has promise in the small-scale, highly-subsampled regime. Table \ref{surveytable} provides a summary of our findings. Finally, we caution again that our tests have been on idealised data, which closely match our model, while data compression and resampling methods require detailed testing.

\section{Summary and Conclusions}

Over the next decade and beyond the size of cosmological data sets, and the potential accuracy and ability to probe new physics, will continue to rise dramatically. To ensure that the expected accuracies are reached, we need to consider the dominant sources of bias and uncertainty in our measurements. 
In this paper we have developed and explored a new framework to study the effect of random errors in the estimation of the data covariance matrix and its inverse, the precision matrix, on cosmological parameter estimation. For multivariate Gaussian data,  the likelihood function depends sensitively on the precision matrix. 

In many areas of cosmology and astrophysics, the data covariance matrix cannot be predicted analytically and we must rely on estimating  it from independent, random realisations of the observations. The simplest estimator of the data covariance matrix  is unbiased, and the uncertainty drops with the inverse square-root of the number of independent samples. More generally, the sample data covariance matrix follows a Wishart distribution.  
In contrast, the simplest estimator for the precision matrix, taking the inverse of the sample data covariance, is biased. We can find an unbiased estimate if the difference between number of realisations, $N_S$, and number of data points $N_D$, is $N_S-N_D > 2$. However, the precision matrix follows an Inverse-Wishart distribution, and its variance can diverge if $N_S-N_D \le 4$.

We have tested and illustrated this behaviour by simulating $10^4$ weak lensing surveys and dividing into 100 groups of $N_S=100$ samples, and shown the mean and variance of the sample data covariance and precision matrix follow the predicted properties of the Wishart and Inverse-Wishart distribution. Future cosmological surveys will have of order $N_D \approx 10^4 - 10^6$ data-points, even with radical compression into power spectra and correlation functions, and so the Wishart bound implies large numbers of realisations will be required. 
 In addition, the only known unbiased estimator of the precision matrix is also the simplest, so other methods will be biased and we may not be able to quantify this bias without comparing with simulations.

The properties of the precision matrix where propagated into the uncertainty in the errors on cosmological parameters, using a Fisher matrix formalism, and we have shown that:
\begin{itemize}
\item The fractional errors  on the variance of  a parameter and the diagonals of the precision matrix are equal.

 \item The fractional error on the parameter variance depends on inverse square-root of  $N_S-N_D-4$, which can diverge when the number of realisations, $N_S$ is equal to $N_D+4$.
 
 \item  The number of realisations needed to reach a given accuracy on parameter errors must be greater than the sum of the number data points, and the inverse of the fractional variance of the parameter variance (equation \ref{eq:Nsim}).
 
 \item The error on the sample data covariance is equal to the precision matrix for small data-sets, while it scales as the inverse-square root of the number of data points for large data sets.

\end{itemize}

If we want to have a $5\%$ accuracy on a parameter error, and the number of data points is $ \ll 100$, we need $ \approx 200$ realisations  and a $10\%$ error on the data covariance matrix. If the data set is greater than 100 points, we will need 
 $N_S \approx N_D$ realisations  and the fractional error on the data covariance matrix is $\approx 10(N_D/200)^{-1/2}$ percent.
We also have shown how the uncertainties propagates into the Figure-of-Merit, and found similar conclusions. 
To attain high-accuracy from large-scale data sets seems to require equally large-numbers of realisations of the survey. We have explored some of the possible alternatives to alleviate this conclusion:
\begin{itemize}
\item {\bf Theoretical Modelling:} If we can accurately model the data covariance we avoid the Wishart bound. 
\item {\bf Shrinkage Estimators:} We reduce the required sample by combining empirical or theoretical estimates of the precision or data covariance matrix with sample estimates.
\item {\bf Data Compression:}  We can reduce the number of data points,  $N_D$, in principle  to the number of parameters. 

\item {\bf Simulations mode resampling:} We can rapidly generate external realisations, but should check for accuracy.
\item {\bf Data Resampling:} The Jackknife resampling method may be useful on small-scales but will be biased. The Bootstrap method can generate large number of samples and so not be biased, but needs further study to avoid other biases.  
\end{itemize}
The combination of theoretical modelling and target shrinkage looks particularly promising and robust, but clearly needs to be developed and tested in more detail for application. 

In summary, many of the details of precision cosmology are still to be worked out. We have identified the estimation of the precision matrix as a key issue, for Gaussian-distributed data, requiring the generation of large numbers of realisations for large-datasets. We have investigated  a number of possible ways forward, although the actual resolution of this issue may require the use of multiple techniques.

\section*{Acknowledgements}
We thank Martin White for useful discussion about the effect of the precision matrix on the parameter error, Peder Norberg for useful discussion about the Jackknife and Bootstrap methods, and Alina Kiessling for encouraging our interest in this problem. We also thank an anonymous referee of useful comments.   BJ thanks the STFC for funding on a Consolidated Grant, while TDK acknowledges the support of a Royal Society University Research Fellowship.

{}

\section*{Appendix A: \\ 
Cosmological Large-Scale Structure Data-sets}

 In general, cosmological data can be in any number of forms, and our analysis is applicable to a wide variety of data. If we are working with pixelised maps then the data are pixel-values and the data covariance matrix is the pixel-covariance matrix. If we have compressed information into two-point power spectra or correlation functions then this forms the data vector, and the data covariance matrix is the field's  four-point function. In this paper we shall assume the data is compressed into the two-point power spectra, although our formulae can be used for pixelised data, correlation functions, or higher-order correlations. Here we outline three basic areas of large-scale structure study, and how their data-vectors are generated.

\subsection*{A1: Galaxy Redshift Surveys}

In galaxy redshift surveys, we can predict the distribution of the matter overdensity field,
 \be
	\delta(\r) = \frac{\rho(\r)-\lgl \rho\rgl}{\lgl \rho\rgl},
\ee
which we can compare with data about the galaxy overdensity, 
 \be
 	\delta_g(\r) =  \frac{n(\r)-\bar{n}(r )}{\bar{n}(r )},
 \ee
were  $n(\r)$ is the galaxy number-distribution and $\bar{n}(r )$ is the survey radial selection function.
The  Fourier transform of the matter overdensity,  $\delta(\k)$,  is 
 \be
 	\delta(\k) = \int \! d^3 \! r  \, \delta(\r) e^{{-i \k.\r}}.
 \ee
  The galaxy distribution can also be expanded in spherical harmonics, $\delta_{\ell m}(z)$, and radial Bessel functions, $\delta_{\ell m n }$ (e.g., Heavens \& Taylor, 1995). 
As we measure galaxy radial positions with redshift, which combines the Hubble expansion with peculiar velocities,  the galaxy distribution  is changed  by redshift-space distortions (Kaiser, 1987; Hamilton, 1998) so that 
in Fourier space, 
\be
	\delta_g^s (\k) = \left( b(k) + f(\Omega_m) \mu_k^2 \right)  \delta(\k) ,
\ee  
where $b(k)$ is a scale-dependent galaxy bias factor, $f(\Omega_m)=d\ln \delta/ d \ln a$ is the growth index. The correlation of the modes of the overdensity field is
 \be
 	\lgl \delta^s_g(\k) \, \delta_g^{s*}\!(\k') \rgl = (2\pi)^3 P^s_{gg}(\k) \delta_D(\k-\k'),
\ee
where $P_{gg}^s(\k)$ is the anisotropic redshift-space galaxy power spectrum. 
We can estimate the anisotropic redshift-space galaxy density power spectrum from 
 \be
 	\widehat{P}^s_{gg}(k,\mu_k) = \frac{1}{N_{\rm modes}}\sum_{k_z} | \delta^s_g(\k)|^2,
 \ee
 where the summation is over the $N_{\rm modes}$  in the $k_z$-direction.
The data is the discretely sampled redshift-space power spectrum;
 \be
 	D_i = \widehat{P}_{gg}^s(\k_i),
 \ee
 where a hat $\,\,\widehat{}\,\,$  indicates the observed estimate of the power.

 \subsection*{A2: CMB}
 
 In  Cosmic Microwave Background experiments  we can define the temperature fluctuations as $\Theta = \Delta T/T$, and the temperature power-spectrum, $C^{TT}_{\ell}$,  is defined by
\be
	\lgl \Theta_{\ell m} \Theta^*_{\ell' m'} \rgl = C^{TT}_\ell \delta_{\ell \ell'}\delta_{mm'}.
\ee
If polarisation data is added to this, in the form of $E-$ and $B-$modes, we can construct 6 power spectra,
\be
	\D = (\widehat{C}^{TT}_\ell, \widehat{C}^{EE}_\ell, \widehat{C}^{BB}_\ell, \widehat{C}^{TE}_\ell, \widehat{C}^{TB}_\ell, \widehat{C}^{EB}_\ell),
\ee
which can form our data-vector. 
 We can estimate these cross-spectra from 
  \be
  	 	\widehat{C}^{XY}_\ell= \frac{1}{(2 \ell+1)}\sum_{m} X_{\ell,m} Y^*_{\ell,m} ,
  \ee
  where $(X,Y)=(\Theta,E,B)$, and we have summed over all azimuthal $m$-modes for each $\ell$.
Again, in practise these power-spectra would be convolved by the survey mask (e.g., Brown, Castro \& Taylor, 2005).

\subsection*{A3: Weak Lensing}

In weak lensing surveys the data can be the estimated shear values,  $\gamma_i(\thetab,z)$, where $i=(1,2)$ are the two orthogonal modes of the shear. The shear-shear covariance matrix, for an unmasked survey, is
\be
	\lgl \gamma_i(\lb, z) \gamma_j^*(\lb', z' )\rgl = (2\pi)^2 C^{\gamma\gamma}_{ij}(\ell,z,'z) \delta_D(\lb-\lb') 
\ee
where $C^{\gamma\gamma}_{ij}(\lb,z,z')$ is the shear power-spectrum. The shear can be decomposed into a potential ($\kappa$, convergence) and curl ($\beta$) part, 
 \be
 	\kappa(\lb,z) + i \beta(\lb,z) = e^{2 i \varphi_\ell}(\gamma_1 + i \gamma_2)(\lb,z),
 \ee
 where $\varphi_\ell$ is the angle between the wavevector,  $\lb$, and an axis of the coordinate system the shear is measured in. This decomposition generates three power spectra, $C^{\kappa\kappa} (\ell,z,z'), C^{\beta\beta} (\ell,z,z'), C^{\kappa\beta} (\ell,z,z')$. In principle, we can also add a magnification field, $\mu$, estimated from the size of galaxy images (e.g., Schmidt et al., 2012;  Casaponsa et al., 2012), yielding the magnification power, $C^{\mu\mu}(\ell,z,z')$. The data, compressed into these power spectra,  is then
 \ba
	\D &=& \big(\widehat{C}^{\mu\mu}(\ell,z,z'), \widehat{C}^{\kappa\kappa}(\ell,z,z'), \widehat{C}^{\beta\beta}(\ell,z,z'),  \nn 
		 & &  \widehat{C}^{\mu\kappa}(\ell,z,z'), \widehat{C}^{\mu\beta}(\ell,z,z'), 
		 \widehat{C}^{\kappa\beta}(\ell,z,z')\big).
\ea
The estimated cross-power spectrum, at two different redshifts, can be estimated by 
  \be
    	\widehat{C}^{XY}_\ell(z,z') = \frac{1}{N_{\rm modes}}\sum_{|{\small \lb}|=\ell} X(\lb,z) Y^*\!(\lb,z'),
  \ee
  where $(X,Y)=(\mu,\kappa,\beta)$, and we have summed over $N_{\rm modes}$ is a shell in $\ell$-space.
 In general, the effects of a survey mask, due to survey geometry and bright stars, will convolve these spectra.  We can also estimate these statistics in real-space, through correlation functions, or other weighted two-point functions such as $M_{ap}$ or COSEBIs (Schneider, et al., 2010;  Asgari, et al., 2012).

\subsection*{A4: Combining data-sets}

Combining data-sets can be achieved by combining the data-vectors of each data-set, and including all the cross-terms between the surveys. To do this we define a vector for field-values; 
 \be
 	\X \! = \! \left(\delta^s_{g, \ell m}(z),  \Theta_{\ell m},E_{\ell m},B_{\ell m}, \mu_{\ell m}(z),\kappa_{\ell m}(z),\beta_{\ell m }(z)\right),
\ee
where we have chosen to expand the lensing shear, magnification and galaxy redshift fields into spherical harmonics for consistency. 
 We then form all of the auto-and cross-spectra of these fields,
 \be
 	C^{X_i X_j}_\ell(z,z') = \frac{1}{2 \ell+1}\sum_{m} X_i(\ell,z) X_j^*(\ell,z'),
 \ee
where we have summed over all azimuthal modes on the sky.
 The data-vector, $\D$, is then the vector of all auto- and cross-spectra.

\section*{Appendix B: \\
Properties of the Wishart and Inverse-Wishart distributions}

Many of the properties of the Wishart and Inverse-Wishart distributions reside in technical mathematical statistics papers and specialist textbooks. Few proofs written for physicists are available, so in this Appendix we present our own derivations of some useful results used in this paper. Most of the results, if not the details,  can be found in Press (1982).

\subsection*{B1. Wishart distribution}

Let $\V$ be a $p \times p$ symmetric matrix so that
 \be
 	\V = \sum_{\alpha=1}^n  \x_\alpha \x_\alpha^t,
 \ee
 where $\x$ is a vector drawn from a multivariate Gaussian distribution, and each of $n$ realisations is sampled independently. The distribution of $\Vb$ is given by (Wishart, 1928)
\be
	p(\Vb|\Sigmab) = c |\Vb|^{(n-p-1)/2} |\Sigmab|^{-n/2}  e^{\small - \half \Tr \Vb \Sigmab^{-1}},
\label{wish}
\ee
where 
\be
	c = \left[ 2^{np/2}   \Gamma_p[n/2]  \right]^{-1} ,
\ee
and $\Sigmab$ sets the scale of the distribution. 
The Multivariate Gamma Function, $\Gamma_p(a)$, is defined as
 \ba
 	\Gamma_p(a) &\equiv & \int_{{\small \X}>0} \! d\X \, |\X|^{a-(p+1)/2} e^{\small - \Tr \X} \\
	& =& 
	 \pi^{p(p-1)/4} \prod_{j=1}^p \Gamma \left(a + \frac{1-j}{2}\right),
 \label{MGF}
 \ea
where $\X$ is a $p \times p$ matrix and the matrix integration is over all positive-definite elements,
\be
	d \X \equiv \prod_{i=1}^p \prod_{j=1}^p d X_{ij},
\ee
or for a symmetric matrix over all non-repeated elements, 
\be
	d \X \equiv \prod_{i=1}^p \prod_{j=1}^i d X_{ij}.
\ee

\subsection*{B2. Derivation of the Wishart distribution}

We can derive the Wishart distribution through its characteristic function, the Fourier transform of the probability distribution function, 
 \be
 	\phi(\Jb) 
			 =  \int d \Vb  p(\Vb|\Sigmab) \, e^{\small i \Tr \Jb \Vb}
			 = \left\lgl e^{\small i \Tr \Jb \Vb} \right\rgl.
 \ee		
Expanding the Fourier exponential in a Taylor series and taking expectations we find, 		
 \be
			\phi(\Jb)
			= 1 + i  \lgl \Tr (\Jb \Vb) \rgl - \half \lgl (\Tr \Jb \Vb)(\Tr \Jb \Vb)\rgl + \cdots .
 \ee
 Using the Gaussian properties of $\x$, the first and second moments of $\Vb$ are
\ba
	\lgl \Vb \rgl &=& \sum_{\alpha=1}^n  \lgl \x_\alpha \x_\alpha^t \rgl 
	= n \Sigmab, \\
	\lgl V_{ij} V_{mn} \rgl &=& 
	\left\lgl \sum_{\alpha=1}^n   x_{i,\alpha} x_{j,\alpha} 
	 \sum_{\beta=1}^n   x_{m,\beta} x_{n,\beta} \right\rgl  \nn
	 &=& 
	n( \Sigma_{im} \Sigma_{jn} + \Sigma_{in} \Sigma_{jm}) + n^2 \Sigma_{ij} \Sigma_{mn}.
\ea
Taking the expectation values, we can re-write the series for the characteristic function in terms of the scale matrix, $\Sigmab$,  as
 \ba
 		\phi(\Jb)
			&=& 1 + i n  \Tr (\Jb \Sigmab) \nn
			& & - 
			\half \left( 2 n \Tr (\Jb \Sigmab \Jb \Sigmab)
			 + 
				n^2 (\Tr \Jb \Sigmab)(\Tr \Jb \Sigmab)\right) \nn
				& & + \cdots.
	\label{cfder}
 \ea
Collecting terms in equation (\ref{cfder}) in powers of $n$ we find
 \ba
 	\phi(\Jb) &=& 1 + n [i \Tr(\Jb \Sigmab)  - \Tr [ (\Jb \Sigmab)^2] + \cdots] \nn
	 & & + \half n^2 [i \Tr(\Jb\Sigmab)-\Tr[(\Jb\Sigmab)^2] + \cdots]^2 + \cdots .
 \ea
Each of the series with factor $n$, $n^2$, etc, can we summed to a logarithm, using the series relation $\ln(1+x) = \sum_{n=0}^\infty(-1)^{n+1} x^n/n$, which yields 
 \ba	 
	\phi(\Jb)	 
		&=& 1- \frac{n}{2} \Tr \ln (\I - 2 i \Jb \Sigmab) + \frac{n^2}{8}[\Tr \ln (\I - 2 i \Jb \Sigmab) ]^2
		\nn
		& & + \cdots .
\ea		
This last series can now be summed, using $e^x = \sum_{n=0}^\infty x^n/n!$, to find 
 \be
	\phi(\Jb)= \exp \left(- \frac{n}{2}\Tr \ln (\I - 2 i \Jb \Sigmab)\right).
 \ee 
Using the matrix identity, $\ln \det \A = \Tr \ln \A$, we find the characteristic function for $\Vb$ is
 \be
 	\phi(\Jb) = |\I - 2 i \Jb \Sigmab |^{-n/2}.
	\label{cf}
 \ee
 
We can show the Wishart distribution has the same characteristic function by direct integration;
 \ba
	  \phi(\Jb) &=& \int d\Vb \, p(\Vb|\Sigmab) \, e^{i {\small \Tr \Jb \Vb}} \nn
		&=& c \, |\Sigmab|^{-n/2} \int d\Vb \,  |\Vb|^{(n-p-1)/2}  
		e^{-  {\small \half \Tr \Vb \Sigmab^{-1} \left[\I -  2i \Jb\Sigmab \right]}}. \nn
 \ea
 It is convenient to define the new matrix variable
 \be
 	\Thetab =  \Sigmab ^{-1} [\I - 2 i \Jb \Sigmab],
 \ee
 and the matrix
 \be
 	\X = \half \Vb \Thetab.
 \ee
The transformation of the matrix volume element is given by 
 \be
  d\Vb = 2^{p(p+1)/2} |\Thetab |^{-(p+1)/2}  d\X .
 \ee
We can now rewrite the characteristic function as
 \ba
 	\phi(\Jb) &=& c \,  |\I - 2 i \Jb \Sigmab|^{-n/2}
	 2^{pn/2} \!\! 
	 \int \! d\X \,  |\X|^{(n-p-1)/2} e^{\small -\Tr \X} \nn
	&=& |\I - 2 i \Jb \Sigmab|^{-n/2}, 
 \ea
 where we have made use of the Multivariate Gamma Function (equation \ref{MGF}) to cancel terms in $c$. Identifying $\phi(\Jb)$ as the characteristic function of $\Vb$ from equation (\ref{cf}), we  confirm that the Wishart distribution, $p(\Vb|\Sigmab)$, is the probability distribution for the matrix $\Vb$. The moments of the Wishart distribution can be found directly from the expansion of the characteristic function in equation (\ref{cfder}).

\subsection*{B3. Inverse-Wishart distribution}

The Inverse-Wishart distribution may be found from the Wishart distribution by a change of variables. We first note that the Jacobian for 
the transformation 
$\U=\V^{-1}$, where $\U$ and $\V$ are $p \times p $ symmetric matrices is (see Appendix C3)
\be
	d\V = |\U|^{-(p+1)} d \U.
 \ee
 If we further define $\G=\Sigmab^{-1}$ then 
 \ba
	p(\U|\G)d\U &=& p(\V|\Sigmab) d\V \nn
			&=&  c \, 
			|\Vb|^{(n-p-1)/2} |\Sigmab|^{-n/2}  e^{\small - \half \Tr \Vb \Sigmab^{-1}} d\V,\nn
			&=&  \left[c\,|\U|^{-(n-p-1)/2} |\G|^{n/2}  e^{\small - \half \Tr \, \U^{-1} \!\G}
			\right]  \nn
			 & & \times  |\U|^{-(p+1)} d\U, \nn
			&=&  c \, |\U|^{-(n+p+1)/2} |\G|^{n/2}  e^{\small - \half \Tr \, \U^{-1} \!\G}   d\U.
\label{invwish}
\ea
Hence, the Inverse-Wishart distribution is
\be
	p(\U|\G) = c\,  |\U|^{-(n+p+1)/2} |\G|^{n/2}  e^{\small - \half \Tr \, \U^{-1} \!\G}.
\ee
We should note that we have assumed the number-of-degrees of freedom, $n$, is the same for the Wishart and Inverse-Wishart, to simplify the derivation. However, the Inverse-Wishart can be parameterized differently, and different authors choose different relations between the Wishart and Inverse-Wishart degrees-of-freedom. If we let $m$ be the number of degrees-of-freedom for the Inverse-Wishart, we have set $m=n$. Other choices commonly used are $m=n+p-1$, $m=n+p+1$, or $m=n-p-1$. For example if we assume $m=n-p-1$ (Press, 1982), we would write the Inverse-Wishart distribution as
\be
	p(\U|\G) = c_0 |\G|^{(n-p-1)/2} |\U|^{-n/2}  e^{\small - \half \Tr \, \G \U^{-1}},
\label{invwish2}
\ee
where 
\be
	c_0 = \left[ 2^{p(n-p-1)/2}   \Gamma_p[(n-p-1)/2]  \right]^{-1} .
\ee
The moments of the Inverse-Wishart can be found by direct integration over the Inverse-Wishart distribution.

\section*{Appendix C:\\
		Change of random matrix variables}

In many cases we want to know how to change random matrix variables, for example in order to carry out matrix integration and to derive the Inverse-Wishart distribution. Here we present some useful matrix transformation relations, without proof. 

\subsection*{C1: Vector transformations}

We first consider vector integration and change of variable. For a $p$-dimensional vector $\x$, the infinitesimal volume-element is
\be
	d\x = d^p \! x = \prod_{i=1}^p d x_i.
\ee
For a vector, $\x$ which is related to the vector $\y$ by the linear transformation,
 \be
 	\y = \A \x,
 \ee
 or with indices $y_i = A_{ij} x_j$,  and the volume element transforms as
\be
	\prod_{i=1}^p d y_i = |\A| \prod_{j=1}^p dx_j
\ee
or equivalently
\be
	d\y = |\A| d\x.
\ee

\subsection*{C2: Transformation of non-symmetric matrices}

If we transform a $p \times p $ matrix $\X$ to the matrix $\Y$,
\be
	\Y = \A \X,
\ee
or with indices, $Y_{ij}=A_{ik} X_{kj}$
the matrix volume-elements for a non-symmetric matrix are
\be
	d\X = \prod_{i=1}^p \left[ \prod_{j=1}^p d X_{ij}\right].
\ee
Each of the sub-vectors transforms as a vector, so that
\be
	\prod_{j=1}^p d Y_{1j}  = |\A| \prod_{j=1}^p d X_{1j} ,
\ee
as each sub-vector has the same determinant of $\A$.
The matrix volume-elements then transform as
\be
	d \Y = |\A|^{p} d\X.
\ee
If we consider now the linear transformation
\be
	\Y = \A \X \B,
\ee
or $Y_{ij} = A_{ik} X_{kl} B_{lj}$, were $\X$ is $p \times q$ and $\B$ is $ q \times q$,  we can apply our transformation rules twice to see that
\be
	d\Y = |\A|^p |\B|^q d\X.
\ee
If $\B=\A^t$, and $q=p$ we find for a general matrix
\be
	d \Y = |\A|^{2p} d\X.
\ee

\subsection*{C3: Transformation of symmetric matrices}

If the matrix $\X$ is a $p \times p$ symmetric matrix, $\X = \X^t$, and we  consider a transformation of the form 
\be
	\Y = \A \X \A^t
\ee
then
\be
	d\Y = |\A|^{p+1} d\X,
\ee
 when $|\A| \ne 0$. In addition, if $\X$ is symmetric and we multiply by a scalar , $a$, so that  
  \be
  	\Y = a \X,
\ee
 then 
  \be
  	d\Y = a^{p(p+1)/2} d\X.
 \ee
  Finally, 
 if we want to transform to the inverse of a symmetric matrix, so that 
  \be
  	\Y = \X^{-1} = \X^{-1} \X \X^{-1},
 \ee 
 then 
  \be
  	d\Y = |\X|^{-(p+1)} d\X.
 \ee

\end{document}